\documentclass[useAMS,usenatbib]{mn2e}
\usepackage[T1]{fontenc}
\usepackage{aecompl}
\usepackage{amsmath,amsfonts,amssymb}
\usepackage[frenchb]{babel}
\NoAutoSpaceBeforeFDP
\usepackage{graphicx}
\usepackage{scrextend}
\usepackage{multirow}
\usepackage{color}

\title[Grain Surface Reactions in Molecular Clouds : The Effect of Cosmic Rays and Quantum Tunneling]{Grain Surface Reactions in Molecular Clouds: The Effect of Cosmic Rays and Quantum Tunneling}
\author[L. Reboussin et al.]{L. Reboussin$^{1,2}$\thanks{E-mail:
laura.reboussin@obs.u-bordeaux1.fr}, V. Wakelam$^{1,2}$, S. Guilloteau$^{1,2}$, F. Hersant$^{1,2}$\\
$^{1}$Univ. Bordeaux, LAB, UMR 5804, F-33270, Floirac, France\\
$^{2}$CNRS, LAB, UMR 5804, F-33270, Floirac, France}
\begin{document}

\date{Received 2013 December 23. Accepted 2014 March 06.}

\pagerange{\pageref{firstpage}--\pageref{lastpage}} \pubyear{2014}

\maketitle

\label{firstpage}

\begin{abstract}

Grain-surface reactions play an essential role in interstellar chemistry, since dust grain catalyses reactions at its surface allowing for the formation of molecules. We used a chemical model in which both gas-phase and grain-surface reactions occur and studied particularly the diffusion mechanisms on the surface of the grains. Surface reactions can occur via thermal hopping when species cross over a potential barrier or via quantum tunneling when species cross through this barrier. We show that the thermal diffusion (hopping) can be much more efficient after a cosmic ray particle collides with a dust grain, heating it to a peak temperature of 70$\,$K. We present here the results of numerical simulations after including the quantum tunneling mechanism for species H, H$_{2}$ and O and considering the effect of cosmic ray particle collision on the surface reactions. As a consequence, the gas-phase and grain-surface abundances are affected and we show that more complex molecules can be formed in molecular clouds.
\end{abstract}
\section{Introduction}
Complex Organic Molecules (COMs) have been detected in warm star-forming regions such as OMC-1 \citep{b27} and Sgr B2 \citep{b28,b29,b30} and are usually associated with warm gas-phase and surface chemistries. COMs are expected to be formed at the surface of the grains during the warm-up phase of star formation when the temperature is about 40 - 50$\,$K \citep{b16}. Under such conditions, heavy radicals can move at the surface of the grains and react without being evaporated. When the temperature gets above 80$\,$K, the COMs are then evaporated in the gas-phase \citep{b34}. Recently, acetaldehyde (CH$_{3}$CHO), dimethyl ether (CH$_{3}$OCH$_{3}$), methyl formate (HCOOCH$_{3}$), methanol (CH$_{3}$OH) and formaldehyde (H$_{2}$CO) have also been observed in the cold pre-stellar cores L1689B \citep{b24} and B1-b \citep{b25}. These detections have revived the question of the formation of these species and their presence in the gas-phase with abundances (relative to the total hydrogen) between $10^{-12}$ to more than $10^{-9}$. \citet{b35} proposed that these complex species are formed at low temperature via gas-phase reactions between precursors, such as CH$_{3}$O, formed on the grain surface and released through reactive desorption. 

At temperature typical of dark clouds or pre-stellar cores (10$\,$K and below), the direct thermal evaporation is only efficient for H and H$_{2}$. Similarly, the diffusion of the species at the surface of the grains, through thermal hopping, is only efficient for atomic hydrogen, so that hydrogen-rich saturated species, such as H$_{2}$O, CH$_{4}$ and NH$_{3}$, are the main constituents of the ices. Current chemical models take into account a number of processes to desorb back into the gas-phase species from the surface. The first one is the desorption induced by stochastic cosmic ray heating \citep{b3}. This process is however only efficient for simple molecules. Another process was introduced by \citet{b11} in which the chemical energy released by exothermic surface reactions contributes to desorb the products. Considering the lack of experimental data on it, the efficiency of the process is not well constrained. \citet{b36} proposed that mantle explosions could be induced by exothermic radical recombination reactions. They concluded that the energy stored by those free radicals being larger than the energy deposits by cosmic rays, the chemical desorption is the dominant non-thermal desorption mechanism in dark clouds. Another desorption mechanism was considered by \citet{b37} in which the energy released by H$_{2}$ formation at the surface of the grains leads to local heating and desorb species back into the gas-phase. However, only weakly bound species (with binding energies less than or equal to that of CO) can be evaporated during this process \citep{b38}. Recent experimental studies by \citet{b39, b43} showed the efficiency of the photodesorption processes. The importance of this mechanism has been underlined by \citet{b45} for the desorption of water at A$_{\text{V}}$ lower than 10 while many observational studies of pre-stellar cores \citep{b46, b44} invoke photodesorption by secondary cosmic ray photons to explain H$_{2}$O gas-phase abundances.

Whatever the desorption mechanism included in the models, the gas-phase abundances reflect the surface abundances. The formation of species at the surface of the grains depends on the diffusion rate of the precursors. This diffusion can be thermal when species migrate from one site to another one by thermal hopping or non-thermal when species cross through a potential barrier by quantum tunneling. Diffusion at the surface of the grains could be faster if tunneling effects are included or if the temperature of the grains is higher. \\

In this paper, we revisit the efficiency of tunneling diffusion based on recent experimental studies of oxygen diffusion \citep{b1}. We also study the effect of cosmic ray impacts which cause a stochastic heating of the dust particles \citep{b2} allowing for surface radicals to diffuse quickly and react to form more complex species. We report here the effect of these two mechanisms on molecular abundances and more specially those of complex organic molecules. Note that we do not distinguish between the different layers of ices in our model. As a consequence, we do not take into account diffusion through the bulk of the ice or any differentiation between the surface and the bulk, which can be an important aspect of surface chemistry as shown by \citet{b42}.\\

The paper is organized as follows. In Section 2 we describe the chemical gas-grain model Nautilus and more particularly the grain-surface reactions. Models predictions are presented in Section 3, while comparisons with observations in two dark clouds (TMC-1 (CP) and L134N) are shown in Section 4. We present our conclusions about this work in the last section.
\section{The Chemical Model: Nautilus}
We used the Nautilus chemical model described in \citet{b9} and \citet {b7} in which the abundance of each species is obtained by solving rate equations for gas-phase and grain-surface chemistries. The kinetic equations describing the formation and destruction of molecules are:
\begin{equation}
\frac{dn_{i}}{dt}=\sum_{l,m}k_{lm}^{i}n_{l}n_{m}-n_{i}\sum_{i\neq l}k_{l}n_{l}+k_{i}^{\text{des}}n_{i}^{s}-k_{i}^{\text{ads}}n_{i},
\end{equation}
\begin{equation}
\frac{dn_{i}^{s}}{dt}=\sum_{l,m}k_{lm}^{i,s}n_{l}^{s}n_{m}^{s}-n_{i}^{s}\sum_{i\neq l}k_{l}^{s}n_{l}^{s}-k_{i}^{\text{des}}n_{i}^{s}+k_{i}^{\text{ads}}n_{i},
\end{equation}
where $n_{i}$ and $n_{i}^{s}$ are, respectively, the gas-phase and surface concentrations of species $i$, $k_{lm}^{i}$ and $k_{lm}^{i,s}$ are the gas-phase and surface reaction rates, and $k_{i}^{\text{des}}$ and $k_{i}^{\text{ads}}$ are the desorption and adsorption rates.
\subsection{Gas-phase chemistry}
The gas-phase bimolecular reaction rates are given by the modified Arrhenius equation as a function of temperature $T$ (in Kelvin):
\begin{equation}
k(T)=\alpha\left(\frac{T}{300}\right)^{\beta}\exp\left(-\frac{\gamma}{T}\right),
\end{equation}
where $\alpha$ is the value of the reaction rate, $\beta$ characterizes the temperature dependence of the rate, and $\gamma$ is the activation barrier (in Kelvin) for exothermic and endothermic reactions with activation energies.\\

Ionization and dissociation rates by 1) direct impact of cosmic ray particles, 2) secondary UV photons induced by cosmic ray/H$_{2}$ interactions, and 3) interstellar FUV photons are also calculated by the model according the following equations \citep{b8}: 
\begin{equation}
k_{\text{CR}}=A_{i}\zeta_{\text{CR}},
\end{equation}
\begin{equation}
k_{\text{FUV}}=A\exp\left(-C A_{\text{V}}\right)\chi,
\end{equation}
where $\zeta_{\text{CR}}$ is the cosmic ray ionization rate (typically $1.3\times10^{-17}\,$s$^{-1}$), $A_{\text{V}}$ is the visual extinction, $\chi$ the FUV flux, $A_{i}$ and $A$ are parameters which represent the rate coefficients and $\exp\left(-C A_{\text{V}}\right)$ takes into account the continuum attenuation from the dust. $\alpha$, $\beta$, $\gamma$, $A_{i}$, $A$ and $C$ are taken from networks (see section 3).

\subsection{Gas-grain interactions}
Gas-phase species can be adsorbed on the grain surface. The rate of adsorption is given by: 
\begin{equation}
k_{\text{ads}}(i)=\sigma_{\text{d}}\langle v(i) \rangle n(i)n_{\text{d}},
\end{equation}
where $\sigma_{\text{d}}$ is the cross section of the grain, $\langle v(i) \rangle$ is the thermal velocity of the species $i$, $n(i)$ its density and $n_{\text{d}}$ the number density of grains.

An adsorbed species can desorb back into the gas-phase. The desorption can be thermal and the corresponding rate is calculated as:
\begin{equation}
k_{\text{des}}(i)=\nu_{0}(i) \exp\left(-\frac{E_{\text{D}}(i)}{T_{\text{g}}}\right),
\end{equation}
where $E_{\text{D}}(i)$ is the desorption energy of the species $i$ (in Kelvin), $T_{\text{g}}$ is the grain temperature assumed to be 10$\,$K and $\nu_{0}(i)$ is the characteristic vibration frequency for the adsorbed species given by:
\begin{equation}
\nu_{0}(i)= \sqrt{\frac{2n_{\text{s}}E_{\text{D}}(i)}{\pi^{2}m(i)}},
\label{nuo}
\end{equation}
where $n_{\text{s}}$ is the surface density of sites ($\sim\,1.5\times10^{15}\,$cm$^{-2}$) and $m(i)$ the mass of the adsorbed species $i$.\\

The gas and dust temperatures are assumed to be the same here, but the grain temperature can be warmer after a grain has been hit by cosmic rays. According to \citet{b2}, a cosmic ray particle (only iron nucleus are considered here) deposits 0.4 MeV on average into dust particles of radius $0.1 \,\mu $m, impulsively heating them to a peak temperature of 70$\,$K. The rate coefficient for non-thermal desorption is given by \citep{b3}: 
\begin{equation}
k_{\text{CR}}=f(70\,\text{K})k_{\text{des}}(i,70\,\text{K}),
\end{equation}
where $f(70$\,$\text{K})$ is the fraction of the time spent by grains at 70$\,$K, defined as the ratio of the time scale for cooling via desorption of volatiles ($\sim10^{-5}\,$s) to the time interval between successive heating to 70$\,$K, which is estimated to be $10^{6}$ years for a cosmic ray ionization rate of $1.3\times10^{-17}\,$s$^{-1}$ \citep{b2}. Then we have: 
\begin{equation}
f(70\,\text{K})=\left(\frac{\zeta_{\text{CR}}}{1.3\times10^{-17}(\text{s}^{-1})}\right)\,3.16\times10^{-19}.
\end{equation}

In addition to the two desorption mechanisms (thermal desorption and cosmic ray desorption) previously mentioned, a non-thermal desorption mechanism via exothermic surface reactions is included in the model. The energy released for each exothermic reaction contributes to desorb the products into the gas-phase. Rice-Ramsperger-Kassel (RRK) theory \citep{b10} is used to obtain the probability of desorption given by: 
\begin{equation}
P=\left[1-\frac{E_{\text{D}}}{E_{\text{reac}}}\right]^{3\text{N}-6},
\end{equation}
where $E_{\text{D}}$ is the desorption energy of the product molecule, $E_{\text{reac}}$ is the energy of formation released and $\text{N}$ ($\text{N}\geqslant2$) is the number of atoms in the product molecule.

To determine the fraction of reactions $f$ for which desorption occurs, we model the competition between the rate of desorption and the rate of energy lost to the grain \citep{b11}: 
\begin{equation}
f=\frac{\nu P}{\nu_{\text{s}}+\nu P}=\frac{cP}{1+cP},
\end{equation}
where $c=\nu/\nu_{\text{s}}=0.01$ is the ratio of the surface molecule bond-frequency to the frequency at which energy is lost (see \citet{b12} for discussion of the value).

\subsection{Grain Surface Reactions}
We only consider Langmuir-Hinshelwood formation mechanism: a species is adsorbed on the surface of an interstellar grain before reacting with another adsorbed reactant. The dust grain is then acting as a third body in chemical reactions. Because of the low grain temperature expected in molecular clouds, we only consider physisorption (by van der Waals force). We assume a sticking probability of 1.0 for neutral atoms that hit the grain. The dust grains, made of amorphous olivine, are assumed to be spherical particles with a radius of 0.1$\,\mu $m, a density of 3$\,$g$\,$ cm$^{-3}$, and we use a dust-to-gas mass ratio of 0.01.\\

Photodissociation by cosmic ray induced UV field and UV photons rates on grain surfaces are calculated according to the same equations as seen in section 2.1 for gas-phase chemistry.\\

The surface reaction rate $R_{ij}$ between species $i$ and $j$ due to diffusion can be expressed as: 
\begin{equation}
R_{ij}=\kappa_{ij}(R_{\text{diff},i}+R_{\text{diff},j})N_{i}N_{j}n_{\text{d}},
\end{equation}
where $N_{i}$ and $N_{j}$ are, respectively, the mean number of molecules of species $i$ and $j$ on a grain, $R_{\text{diff}}$ is the diffusion rate defined as the inverse of the diffusion time $t_{\text{diff}}$ and $\kappa_{ij}$ is the probability for the reaction to occur. For a reaction without activation energy, the probability for the reaction to happen is unity. For a reaction with activation energy $E_{\text{a}}$, the probability is given by: 
\begin{equation}
\kappa_{ij}=\alpha \exp\left(-\frac{E_{\text{a}}}{T_{\text{g}}}\right).
\end{equation}

Two diffusion mechanisms are included in the model: diffusion by thermal hopping when species cross over a potential barrier and diffusion by quantum tunneling when species cross through the potential barrier.
\subsubsection{Diffusion by thermal hopping}
In order to react, the adsorbed species requires mobility. The time scale for an adsorbed species to migrate from one site to another one via thermal hopping is given by the equation:
\begin{equation}
t_{\text{hop}}=\nu_{0}^{-1}\exp \left(\frac{E_{\text{b}}}{T_{\text{g}}}\right),
\label{thop}
\end{equation}
where $\nu_{0}^{-1}$ is from eq. \eqref{nuo} and $E_{\text{b}}$ is the energy barrier (in kelvin) between two adjacent sites, which is a fraction of the binding energy $E_{\text{D}}$. Different values have been assigned to $E_{\text{b}}$ in previous models: from 0.3 $E_{\text{D}}$ \citep{b13,b14,b9} to 0.77 $E_{\text{D}}$ \citep{b15}. In our work, we adopt the estimate of \citet{b16} in which $E_{\text{b}}$ is taken as half of the binding energy.\\

The diffusion time, which is the time for an adsorbed species to scan the entire surface of the grain, is given by:
\begin{equation}
t_{\text{diff}}=N_{\text{S}}t_{\text{hop}},
\end{equation}
where $N_{\text{S}}=10^{6}$ is the total number of surface sites on a 0.1 $\,\mu $m grain.
\subsubsection{Diffusion by quantum tunneling}
The diffusion via thermal hopping requires sufficient energy to overcome the energy barrier between two adjacent sites. Barrier penetration by quantum tunneling can then be more effective for some species, particularly the lighter ones. The time scale $t_{\text{q}}$ for species to migrate to an adjacent surface site through a rectangular barrier via quantum tunneling \citep{b9} is:
\begin{equation}
t_{\text{q}}=\nu_{0}^{-1}\exp\left(\frac{2a}{\hbar}\sqrt{2mE_{\text{b}}}\right).
\label{tq}
\end{equation}
where $a$ is the barrier thickness estimated to be 1 angstr\"{o}m.\\

Diffusion by quantum tunneling depends on the particle mass: tunneling reaction involving a species with a lighter mass is faster than with a heavier mass. For $E_{\text{b}}=225\,$K, the quantum diffusion time for an H atom over an entire grain obtained from equation \eqref{tq} is $\sim1.3\times10^{-4}\,$s, whereas the thermal hopping diffusion time obtained from equation \eqref{thop} is $\sim\,1.7\times10^{3}\,$s. For H$_{2}$, considering $E_{\text{b}}=250\,$K, equations \eqref{tq} and \eqref{thop} give, respectively, the values $\sim3.5\times10^{-3}\,$s and $\sim\,2.8\times10^{4}\,$s.\\

Therefore, because of their low mass and high mobility, we considered surface migration by quantum tunneling for light species H and H$_{2}$. Quantum tunneling for heavier species is only considered for O here, since \citet{b1} presented that, despite its high mass, quantum diffusion may be efficient for oxygen. Their experiments, performed  on physisorbed O atoms via the study of O$_{3}$ formation, showed that oxygen atoms diffusion is governed by quantum tunneling up to 20$\,$K. Furthermore, \citet{b41} also studied theoretically the O + CO reaction and showed that tunneling strongly increases the reaction rates at low temperature (10 - 20$\,$K).
\subsection{Cosmic Ray Induced Diffusion rate (CRID)}
Diffusion on dust particles strongly depends on surface temperature. A higher grain temperature leads to a higher mobility of species on the surface. When a cosmic ray particle collides with a dust grain, it deposits its energy, heating the grain to a peak temperature of 70$\,$K \citep{b2}. The dust grain temperature is then no longer 10$\,$K but seven times warmer during the fraction of the time $f(70$\,$K)$. It is the effect of this stochastic heating on the diffusion that we are studying in this paper. This heating increases the mobility of radicals on the surface of the grain and species will be able to quickly scan the surface since the time scale for species to migrate to an adjacent site will be much faster:
\begin{equation}
t_{\text{hop(CR)}}=\nu_{0}^{-1}\exp \left(\frac{E_{\text{b}}}{T_{\text{CR}}}\right),
\label{tcr}
\end{equation}
where $T_{\text{CR}}$ is the temporary peak temperature of the grain after heating.\\

Therefore, the recombination of radicals on the surface of the grain will be more efficient. The surface reaction rate $R_{\text{CR}}$ between species $i$ and $j$ due to the diffusivity by cosmic rays is then given by:
\begin{equation}
R_{\text{CR}}=f(70\,\text{K})\kappa_{ij(70\,\text{K})}(R_{\text{diff},i}+R_{\text{diff},j})N_{i}N_{j}n_{\text{d}},
\end{equation}
where $R_{\text{diff}}=(N_{\text{S}}t_{\text{hop(CR)}})^{-1}$ is the diffusion rate for a grain temperature of 70$\,$K.
\section{Results of the numerical simulations}
The chemical network used for this work contains 8624 reactions: 6844 are gas-phase reactions and 1780 are grain-surface and gas-grain interactions. The model follows the chemistry of 703 species (atoms, radicals, ions and molecules): 504 are gas-phase species and 199 are species on grains. The surface network is based on \citet{b11} whereas the gas-phase network is based on kida.uva.2011 and that we have updated based on \citet{b32} and \citet{b31}. The full network will be available on the KIDA (KInetic Database for Astrochemistry) website \footnote{http://kida.obs.u-bordeaux1.fr/models}.\\ For the simulations, we consider a typical dense cloud as our nominal cloud: gas density of $n=n(H)+2n(H_{2})=2\times10^{4}\,$cm$^{-3}$, temperature of 10$\,$K, visual extinction of 10, cosmic ray ionization rate of $1.3\times10^{-17}\,$s$^{-1}$ and the elemental abundances as listed in Table 1. We run different models as described in Table 2.
\begin{table}
 \centering
 \begin{minipage}{8cm}
  \caption{Elemental Abundances.}
  \begin{tabular}{ll}
  \hline
   Element & Abundance (/H)\\
\hline
H$_{2}$&0.5\\
He&{9$\times10^{-2}$
   \footnote{see discussion in \citet{b4}.}}\\
N&{6.2$\times10^{-5}$
 \footnote{\label{bb}\citet{b5}.}}\\
O&{1.4$\times10^{-4}$
   \footnote{see discussion in \citet{b6}.}}\\
C$^{+}$&{1.7$\times10^{-4}$
   \footref{bb}}\\
S$^{+}$&{8$\times10^{-9}$
    \footnote{\label{dd}Low metal-elemental abundances \citep{b33}.}}\\
Si$^{+}$&{8$\times10^{-9}$
   \footref{dd}}\\
Fe$^{+}$&{3$\times10^{-9}$
   \footref{dd}}\\
Na$^{+}$&{2$\times10^{-9}$
   \footref{dd}}\\
Mg$^{+}$&{7$\times10^{-9}$
   \footref{dd}}\\
P$^{+}$&{2$\times10^{-10}$
   \footref{dd}}\\
Cl$^{+}$&{1$\times10^{-9}$
   \footref{dd}}\\
\hline
\label{table1}
\end{tabular}
\end{minipage}
\end{table}

\begin{table}
 \centering
 \begin{minipage}{5cm}
 \caption{Models description.}
\begin{tabular}{lllll}
\hline
\multirow{2}{*} {\bf \small{Model A}} & \small{thermal hopping} \\
     &  \small{C/O=1.2} \\ \hline
\multirow{3}{*} {\bf \small{Model B}} & \small{thermal hopping} \\
     & \small{quantum tunneling} \\
     &  \small{C/O=1.2} \\ \hline
\multirow{3}{*} {\bf \small{Model C}} & \small{thermal hopping} \\
     & \small{CRID} \\
     &  \small{C/O=1.2} \\ \hline
\multirow{4}{*} {\bf \small{Model D}} & \small{thermal hopping} \\
     & \small{quantum tunneling}\\
     & \small{CRID} \\
     &  \small{C/O=1.2} \\ \hline     
\multirow{2}{*} {\bf \small{Model E}} & \small{thermal hopping} \\
     &  \small{C/O=0.7} \\ \hline
 \multirow{3}{*} {\bf \small{Model F}} & \small{thermal hopping} \\
     & \small{quantum tunneling}\\
     &  \small{C/O=0.7} \\ \hline
  \multirow{3}{*} {\bf \small{Model G}} & \small{thermal hopping} \\
     & \small{quantum tunneling: 2$^{nd}$ method} \\
     &  \small{C/O=1.2} \\ \hline
 \end{tabular}
 \end{minipage}
 \end{table}
In the following sections, we present the results obtained considering the new mechanisms. We first show the computed abundances after including in the model the quantum tunneling diffusion for H, H$_{2}$ and O (see section 2.3.2). Secondly, we study the effects of cosmic ray on the diffusion rates (see section 2.4) and its sensitivity to the visual extinction, the dust temperature and the cosmic ray ionization rate.
\subsection{The effects of quantum tunneling}
\begin{figure}
\centering
\includegraphics[scale=0.3]{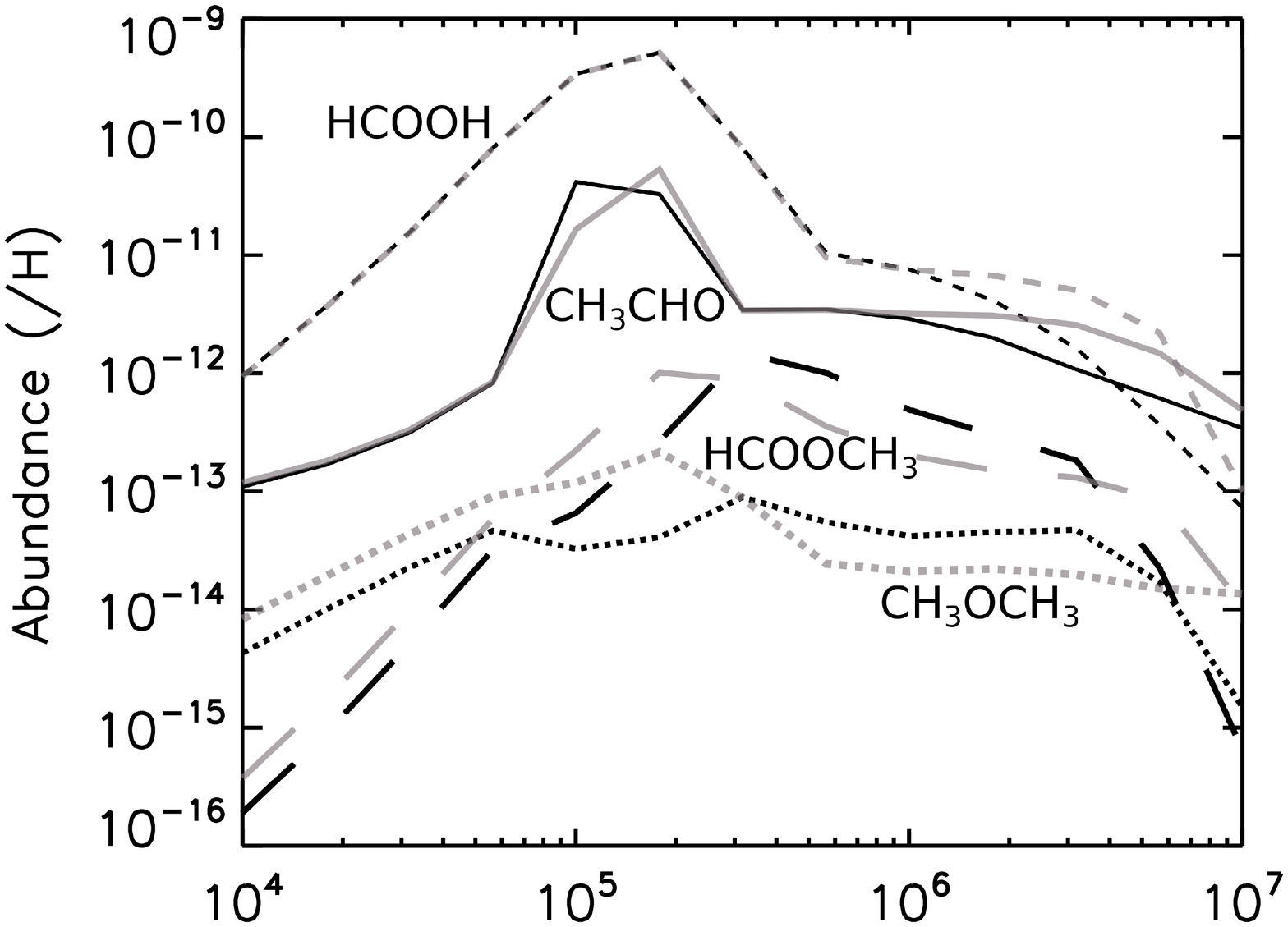}
\includegraphics[scale=0.3]{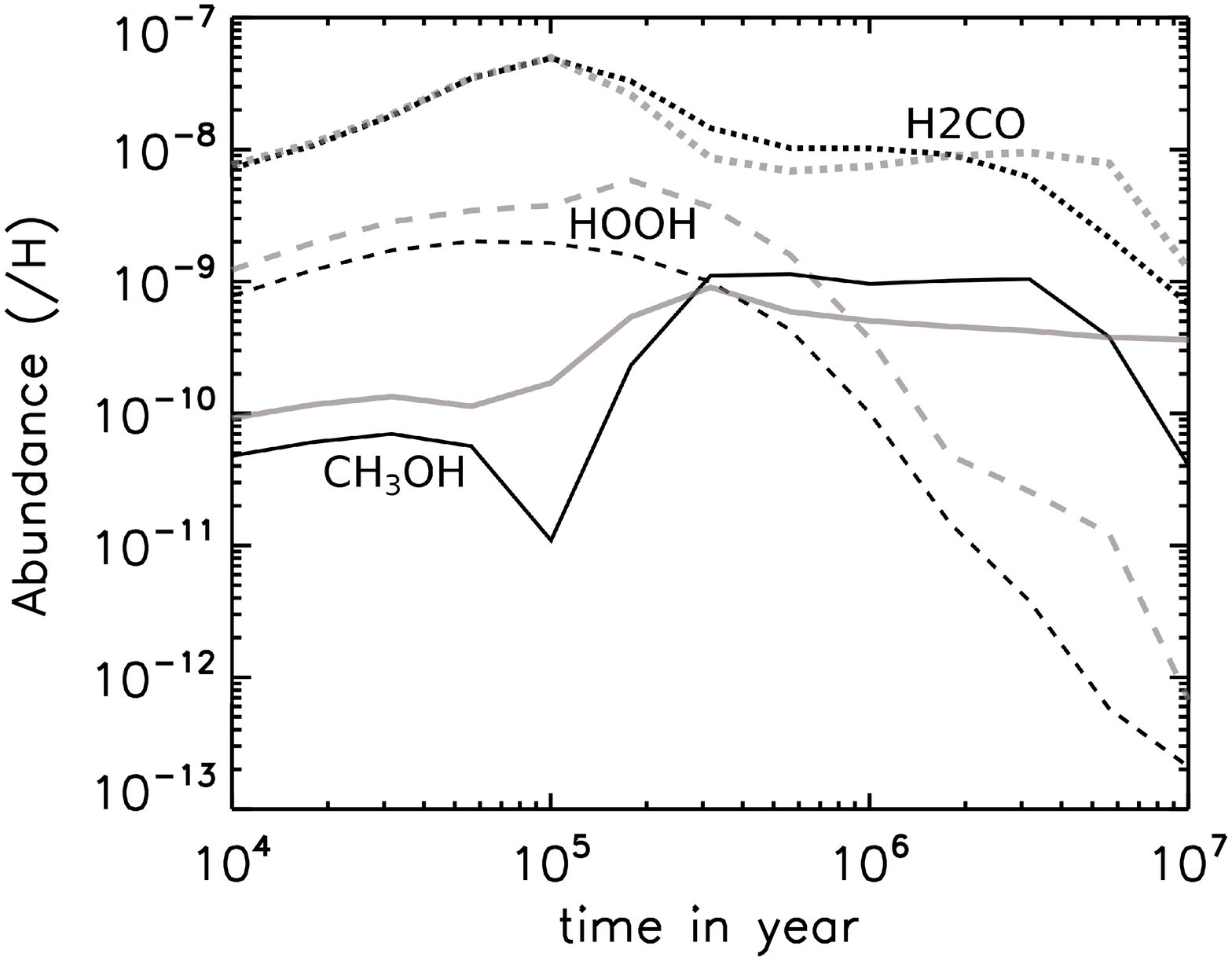}
 \caption{Gas-phase abundances of a selection of species relative to the total hydrogen as a function of time. Black lines represent the results obtained with Model A (without diffusion by tunneling) and grey lines those obtained with Model B (including diffusion by tunneling for species H, H$_{2}$ and O).}
 \label{gasph1}
\end{figure}
\begin{figure}
\centering
\includegraphics[scale=0.3]{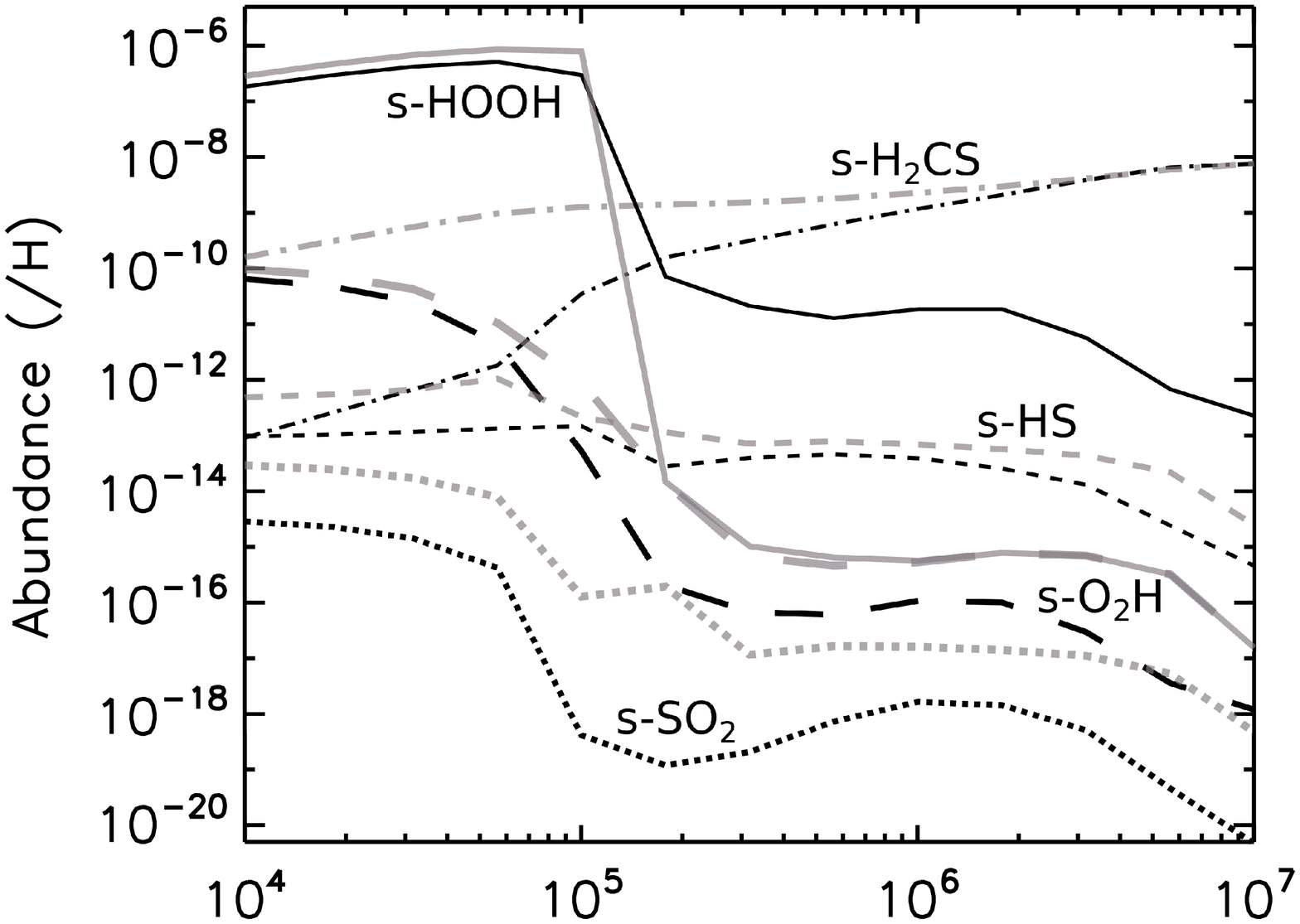}
\includegraphics[scale=0.3]{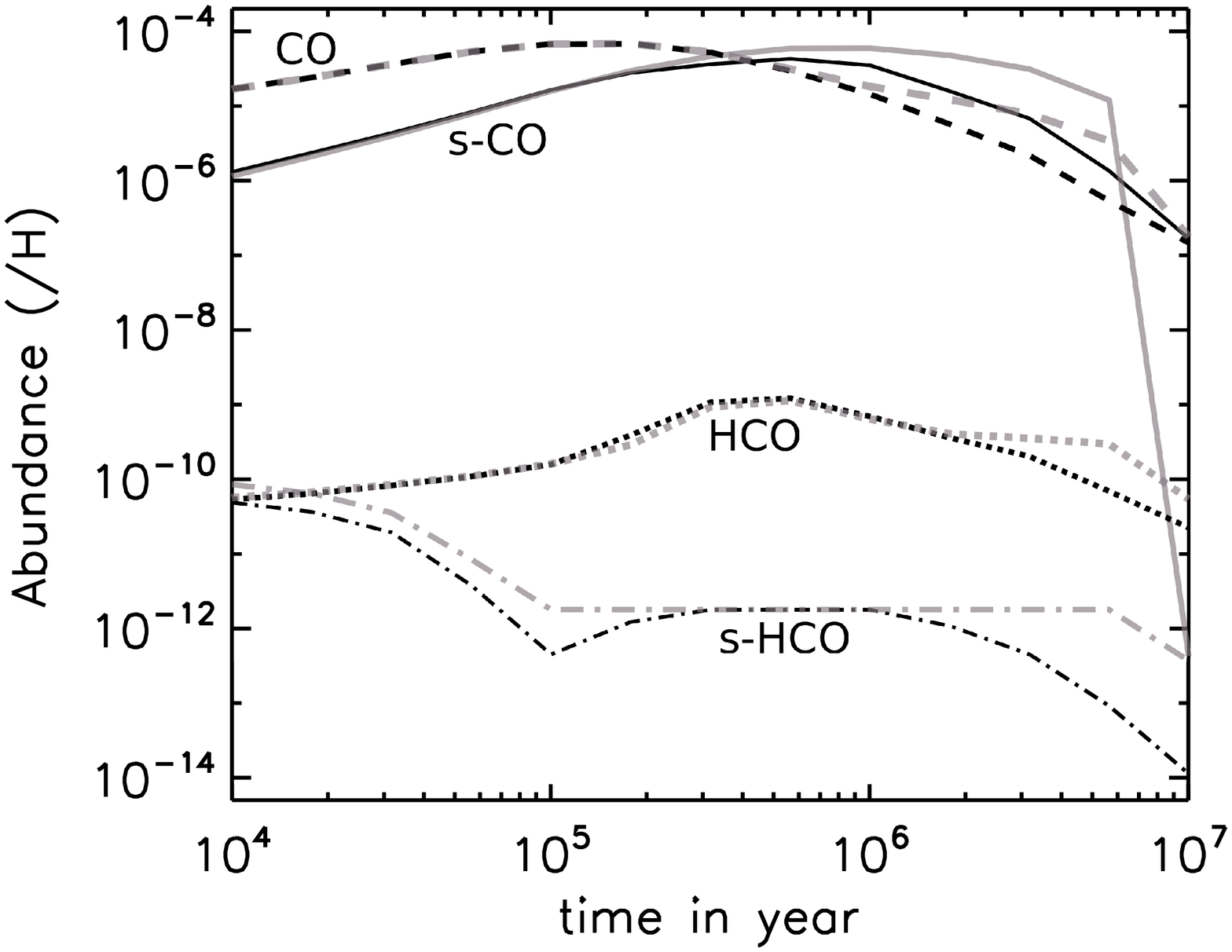}
\caption{Species abundances as a function of time (s- refers for surface species). Results obtained with Model A are shown in black lines while those obtained with Model B are shown in grey lines.}
 \label{graph2}
\end{figure}
We consider a first model which only takes into account the diffusion via thermal hopping (hereafter Model A) and a second model in which diffusion via quantum tunneling for H, H$_{2}$ and O has been added (hereafter Model B). In this last one, the effect described in section 2.4 (CRID) is not included. The physical parameters are the ones of our nominal cloud. Fig. \ref{gasph1} shows the computed gas-phase abundances for a selection of complex molecules as a function of time (black lines for Model A and grey lines for Model B). These species were chosen because they present particular sensitivity to the studied process. We can see that the abundances of HCOOCH$_{3}$, CH$_{3}$OCH$_{3}$ and CH$_{3}$OH exhibit similar sensitivities to the studied effect. The quantum tunneling first produces larger abundances for these three species before $3\times10^{5}$ years. Their abundances are then smaller by a factor of $\sim$ 5 until about $5\times10^{6}$ years. After this time, the tunneling prevents their abundances from dropping as it happens in Model A. As a consequence, these molecular abundances are larger by a factor of $\sim$ 10 after $10^{7}$ years in Model B compared to Model A. The other molecules, HCOOH, HOOH, CH$_{3}$CHO and H$_{2}$CO, are mostly increased, in particular HOOH. They present an increase by a factor of 3 to 10 after $10^{7}$ years. The corresponding species at the surface of the grains present similar sensitivities to the tunneling effect. Considering H, H$_{2}$ and O quantum tunneling diffusion enhances the formation of oxygen-bearing and hydrogen-bearing species, suggesting that quantum tunneling may be able to improve formation of more complex species in cold regions.

Simpler species are also affected by the quantum tunneling mechanism. Fig. \ref{graph2} (top) represents the computed grain-surface abundances for a selection of such species as a function of time. Species abundances are heavily enhanced with Model B, up to a factor of $\sim$ 1000 for s-H$_{2}$CS. Only s-HOOH sees its abundance decreasing after $10^{5}$ years due to the fact that the reaction s-H + s-HOOH $\rightarrow$ s-O$_{2}$H + s-H$_{2}$ is much more effective (its reaction rate is $10^{5}$ times higher) when we take into account the tunnel effect. H$_{2}$CS, HS, O$_{2}$H and SO$_{2}$ gas-phase abundances are enhanced with Model B but much less than their analogous surface species.
As we can see on Fig. \ref{graph2} (bottom), CO abundance is affected as well. The diffusion reaction between s-O and s-C being 1000 times more efficient with Model B, its abundance increases like that of s-CO until $6\times10^{6}$ years. After this time, s-CO abundance decreases because of the fact that s-CO and s-H recombine efficiently to form HCO and s-HCO. Most of the reactions with s-H, s-H$_{2}$ and s-O are much more effective with Model B (with a reaction rate 10 to $10^{5}$ times higher) which affects both grain-surface and gas-phase abundances.
\subsubsection{Sensitivity to the visual extinction, the C/O elemental ratio and the dust temperature}
\begin{figure}
\centering
\includegraphics[scale=0.3]{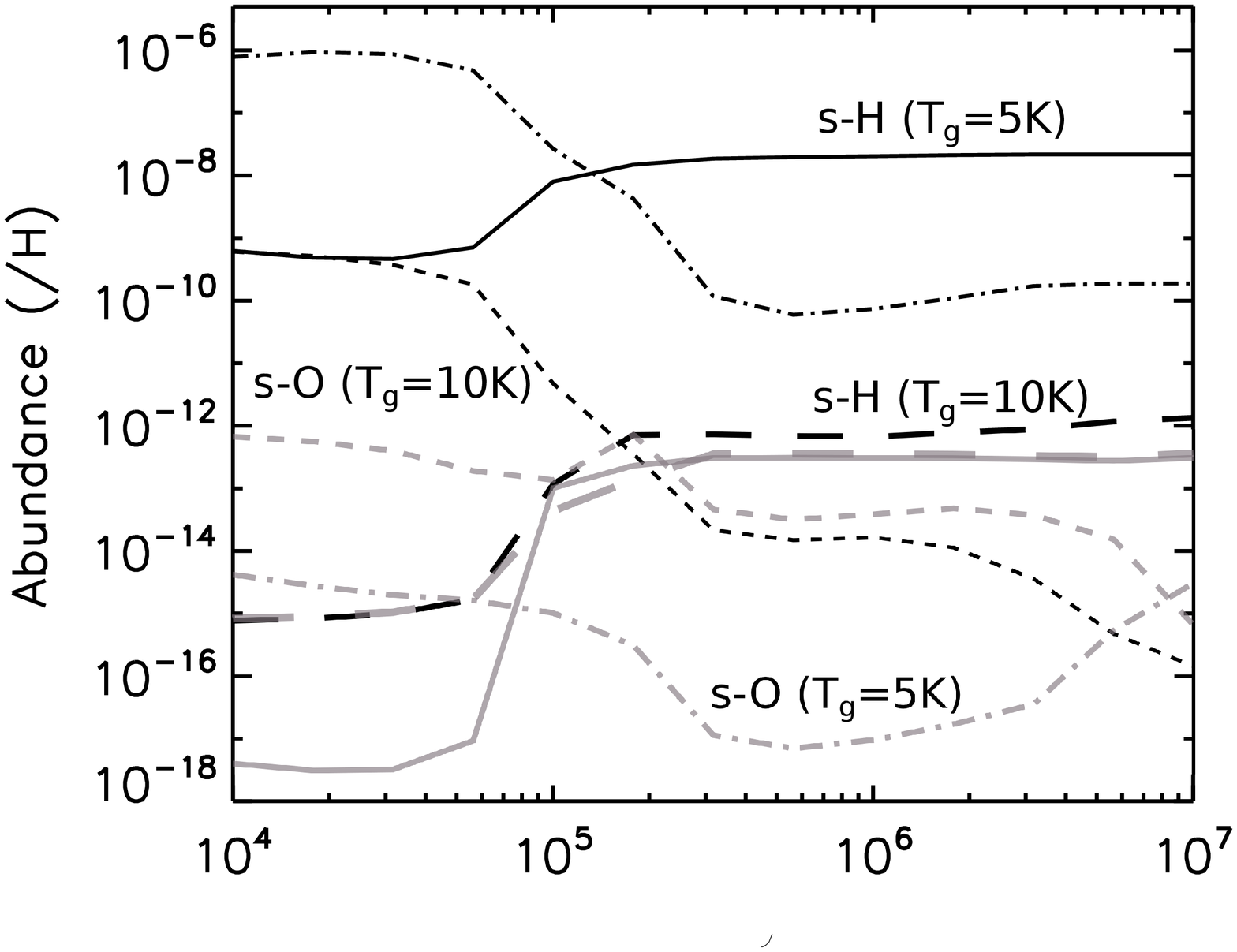}
\includegraphics[scale=0.3]{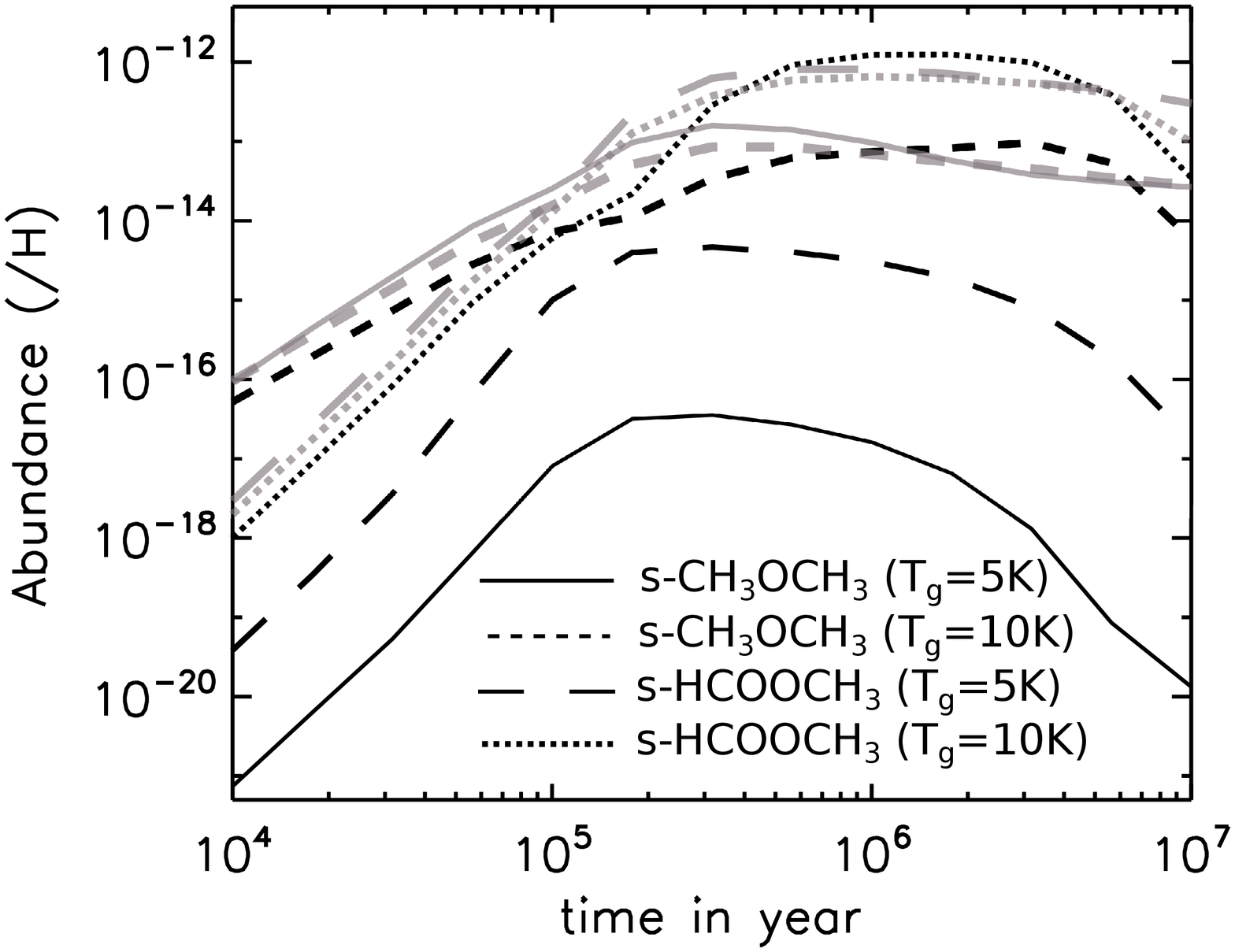}
\caption{Grain-surface abundances as a function of time for two different grain temperatures (5 and 10$\,$K). Results obtained with Model A are shown in black lines while those obtained with Model B are shown in grey lines.}
\label{graph4}
\end{figure}
To see the effect of the cloud visual extinction on the importance of quantum tunneling, we run our models A and B with visual extinctions between 2 and 5.  Our results show that the importance of tunneling versus thermal hopping does not depend much on the visual extinction. The effect of photodissociation process as a function of the visual extinction is much more important than the quantum tunneling mechanism itself, meaning that the formation of species enhanced by tunneling effect does not prevent for the decrease in complex molecular abundances at low A$_{\text{V}}$.

In the simulations presented up to now, we have used a C/O elemental ratio of 1.2. Since this parameter is not well known, we also run our models with a higher elemental abundance for the oxygen ($2.4\times10^{-4}$ (/H)). With this ratio of 0.7, our general conclusions of the paper are not changed. In both cases (smaller A$_{\text{V}}$ or C/O), the factor of difference between abundances computed with or without diffusion by tunneling effect is almost the same whatever the model parameters.

We also studied the effect of dust temperature on the importance of quantum tunneling mechanism versus thermal hopping. In that case, there is a direct effect of this parameter since the rate of the classical diffusion increases with the temperature (see eq. \eqref{thop}). Fig. \ref{graph4} shows the computed abundances of a selection of species for different grain temperatures. We can see that tunnel effect is much more important when the dust temperature is lower. H and O abundances decrease more when the dust temperature is lower, whereas those of more complex species increase more strongly. The lower the temperature, the more diffusion via quantum tunneling predominates over diffusion via thermal hopping. At lower temperature, quantum tunneling will have a very strong effect for the formation of complex species.
\subsubsection{The tunnel effect:  an alternative method}
\begin{figure}
\centering
\includegraphics[scale=0.3]{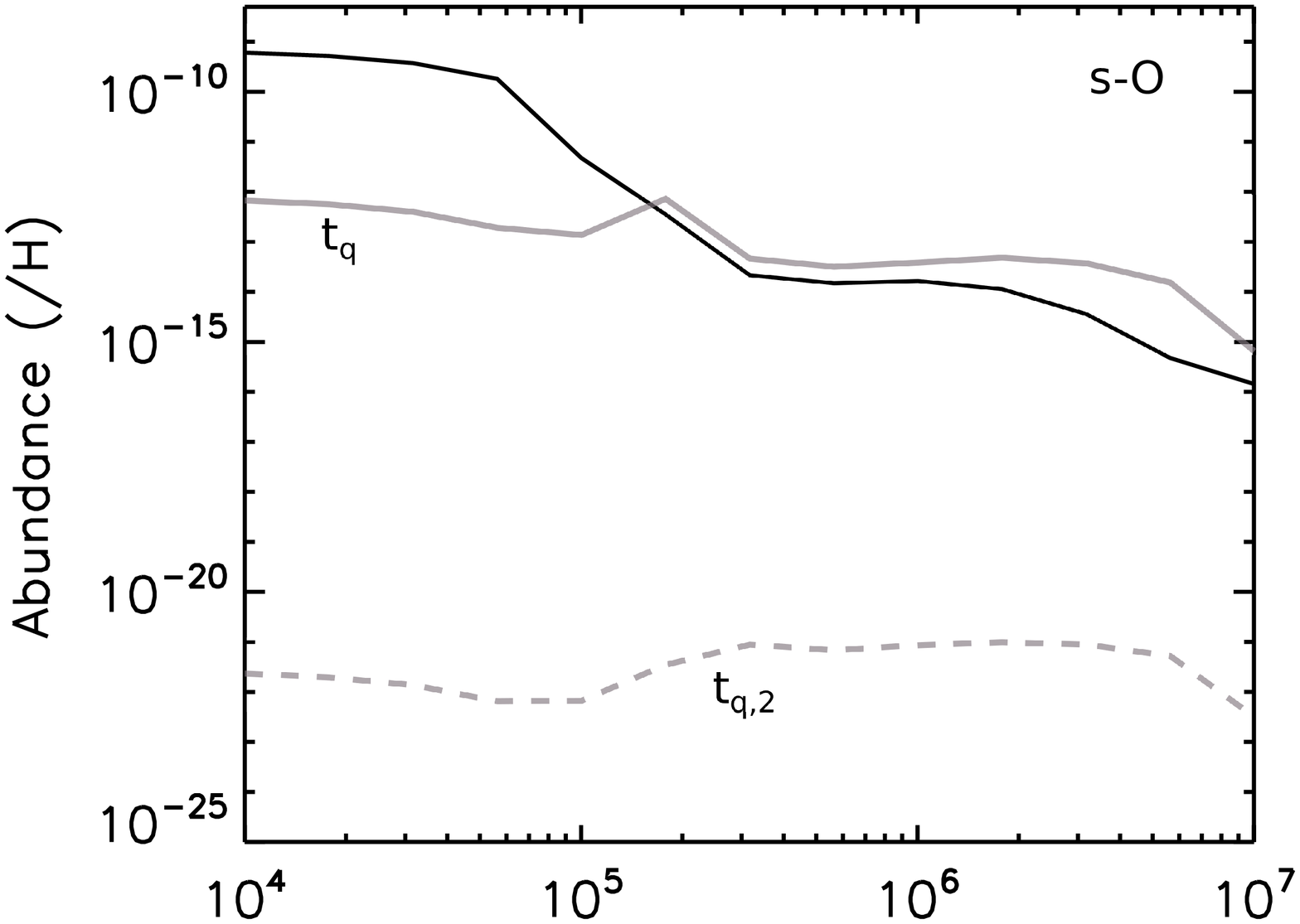}
\includegraphics[scale=0.3]{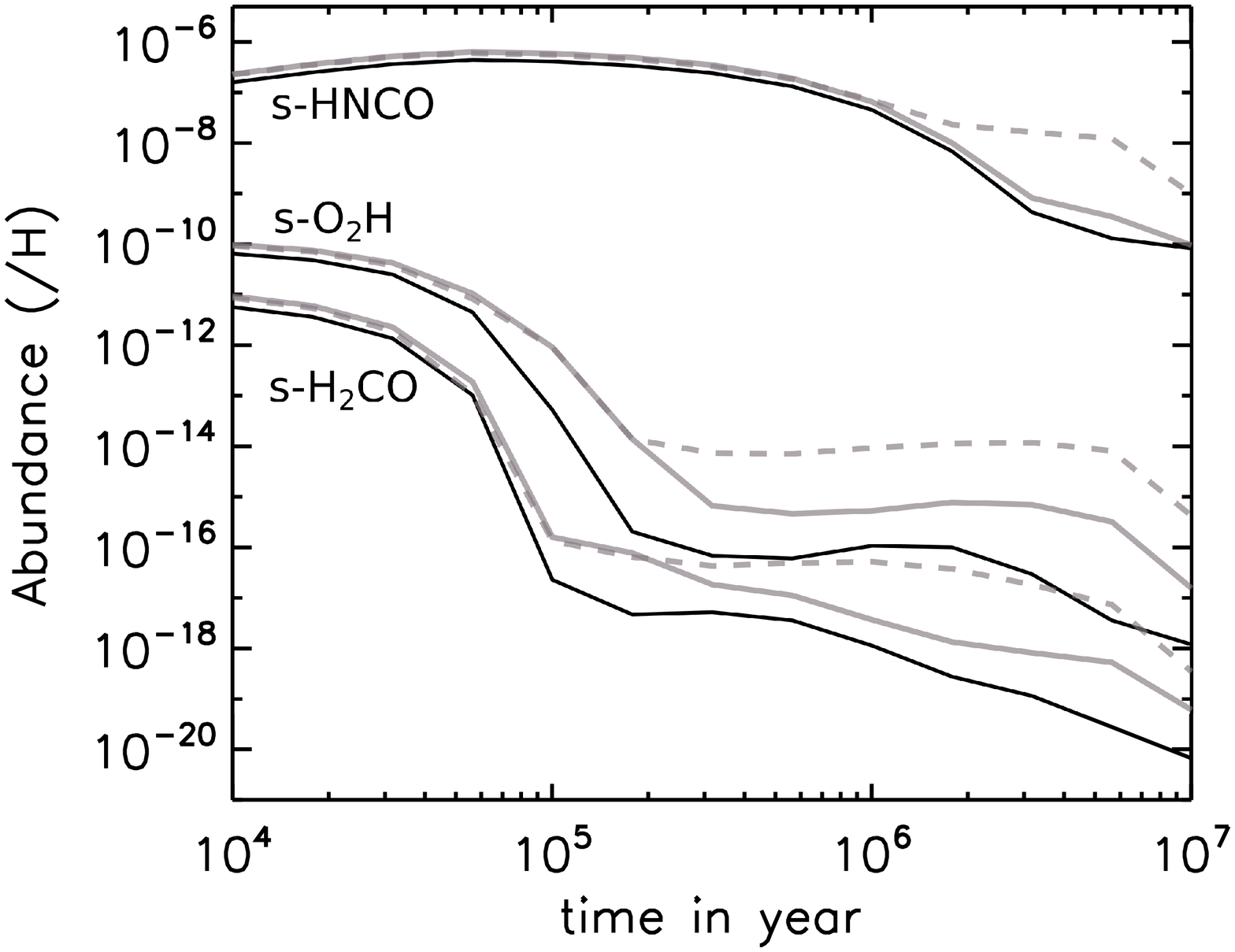}
\caption{Abundances of some species as a function of time using the two different methods for the calculation of the diffusion time by quantum tunneling. Solid black lines represent the results obtained with Model A (without diffusion by tunneling), solid grey lines those obtained with Model B (diffusion by tunneling using equation \eqref{tq}) and dashed grey lines those obtained with Model G (diffusion by tunneling using the equation \eqref{tq2}). }
\label{rq.eps}
\end{figure}
\begin{table}
 \begin{minipage}{7.5cm}
  \caption{Bandwidths and diffusion times for surface states of adsorbed atoms.}
  \begin{tabular}{llll}
  \hline
Atom/surface&$\Delta E_{\text{b}}$ (K)&$N_{\text{S}}$$t_{\text{q,2}}$ (s)&$N_{\text{S}}$$t_{\text{q}}$ (s)\\
 \hline
H/H$_{2}$O&30&$10^{-6}$&$1.3\times10^{-4}$\\
 H$_{2}$/H$_{2}$O&7&$4.4\times10^{-6}$&$3.5\times10^{-3}$\\
 O/H$_{2}$O&$7.5\times10^{-4}$&$4.1\times10^{-2}$&$10^{8}$\\
  \hline
\label{table3}
\end{tabular}
$\Delta E_{\text{b}}$ from \citet{b13}.
\end{minipage}
\end{table}
The diffusion time by quantum tunneling being sensitive to small uncertainties, especially the barrier thickness $a$, which is poorly constrained, quantum mechanical theoretical studies have been made by \citet{b17} using delocalized wavefunctions. They showed that the time scale to migrate to an adjacent potential well by quantum tunneling is given by the Heisenberg uncertainty principle: 
\begin{equation}
t_{\text{q,2}}=\frac{4\hbar}{\Delta E_{\text{b}}},
\label{tq2}
\end{equation}
where $\Delta E_{\text{b}}$ is the gap between the energy bands corresponding to the fundamental and first excited state \citep{b21}. Values of $\Delta E_{\text{b}}$ found in the literature are listed in Table 3.\\

Diffusion time is much faster by using the equation \eqref{tq2}, it is increased by a factor 100 and 1000 for H and H$_{2}$ respectively and even by a factor of $10^{10}$ for O compared to the one obtained with equation \eqref{tq}. The consequences on the abundances concern mostly the O-bearing and H-bearing species. As we can see in Fig. \ref{rq.eps}, the oxygen is much more consumed with this alternative method for the calculation of the diffusion time, since its abundance decreases by a factor of $\sim$ $10^{6}$ after $10^{7}$ years. As a consequence, abundances of O$_{2}$H, HOOH, O$_{3}$, HNCO, HC$_{2}$O, s-HC$_{2}$O, s-CCO, s-CO$_{2}$, s-HNCO, s-O$_{2}$H, s-O$_{3}$, s-OCN and s-SO$_{2}$ are increased by a factor of $\sim$ 10. However, more complex organic molecules are not affected. This formalism for the time scale of diffusion on the surface is much more efficient than the one presented in section 2.3.2. This method however strongly depends on experimental support for the value of $\Delta E_{\text{b}}$ and no new experiment has been conducted since the 1970s. In addition, there are inconsistencies in the literature between \citet{b13} and \citet{b17}. Considering the possible effect of this mechanism, new experimental studies could improve strongly our view of the interstellar chemical modeling.
\subsection{The effects of cosmic ray induced diffusion}
In order to study the effect of cosmic ray impact on surface species diffusion, we considered two models. In the first one, we consider that the species diffuse and react at 10$\,$K only (Model A previously described) whereas in the second model, we add the diffusion of species due to stochastic cosmic ray heating as described in section 2.4 (hereafter Model C). In both cases, we have not included any tunneling effects for diffusion in the models. We first run the two models for the physical parameters of our nominal cloud. For these conditions, we found no effect of the CRID. 
\subsubsection{Sensitivity to visual extinction}
\begin{figure}
\begin{center}
\includegraphics[scale=0.3]{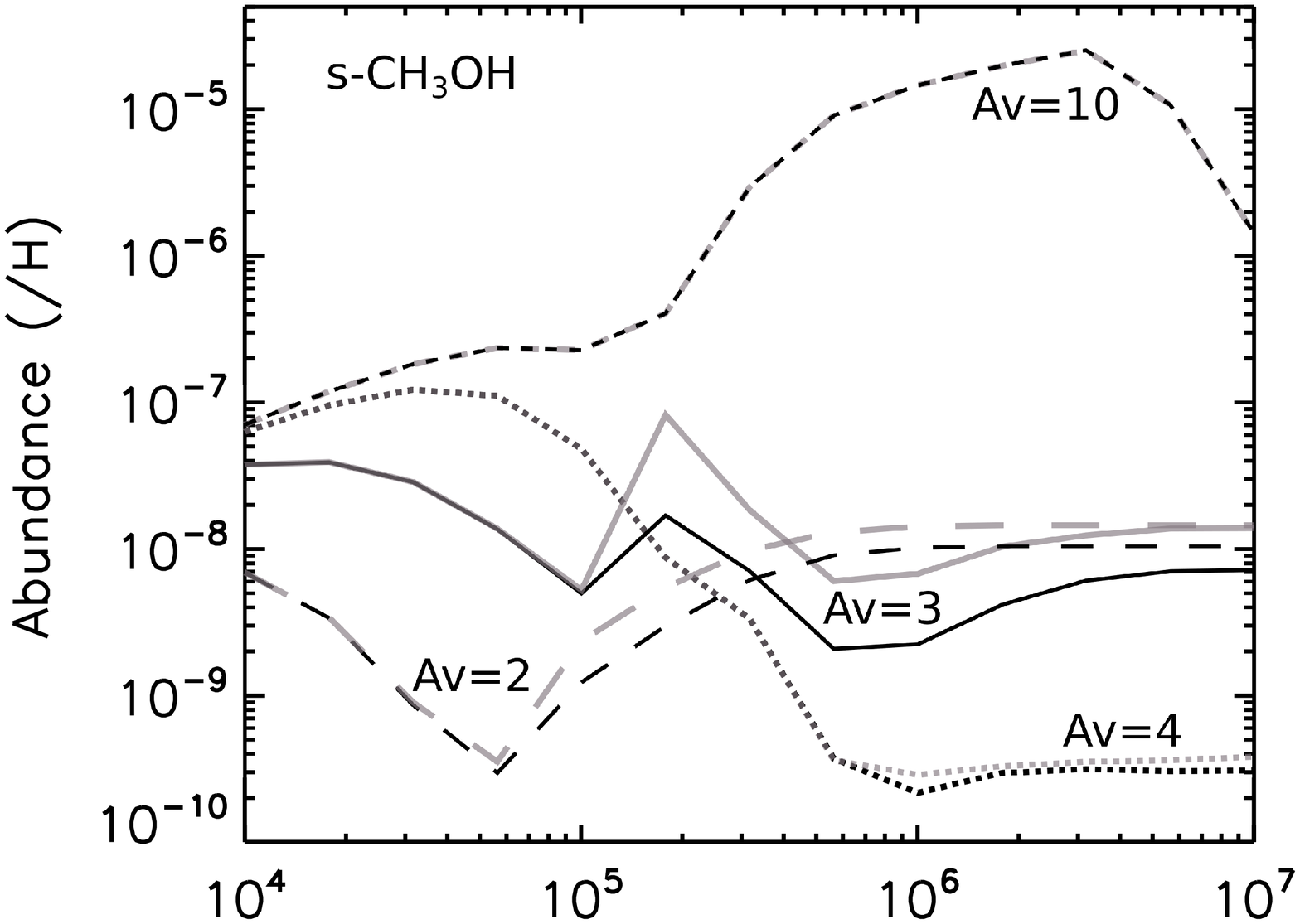}
\includegraphics[scale=0.3]{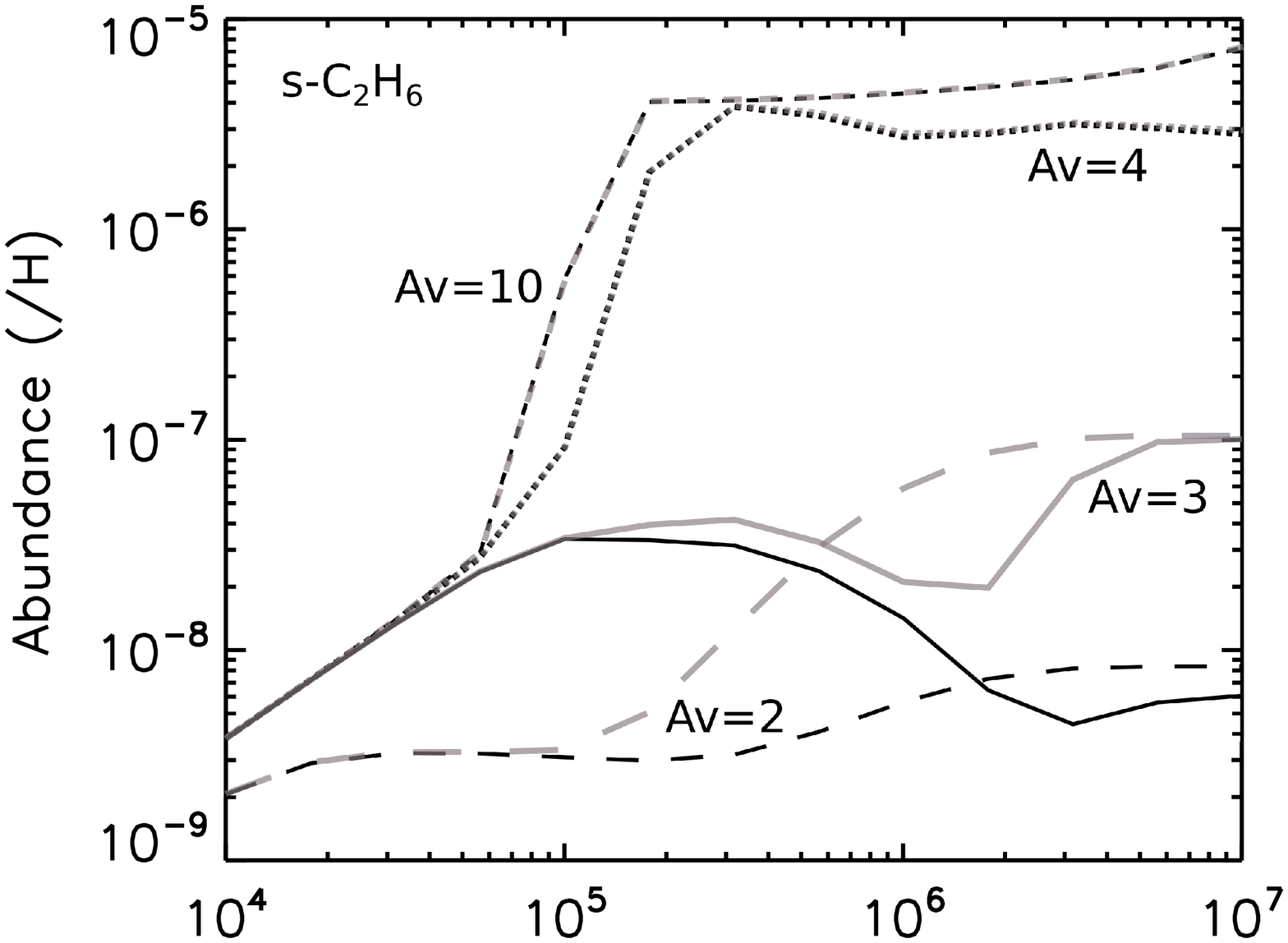}
\includegraphics[scale=0.3]{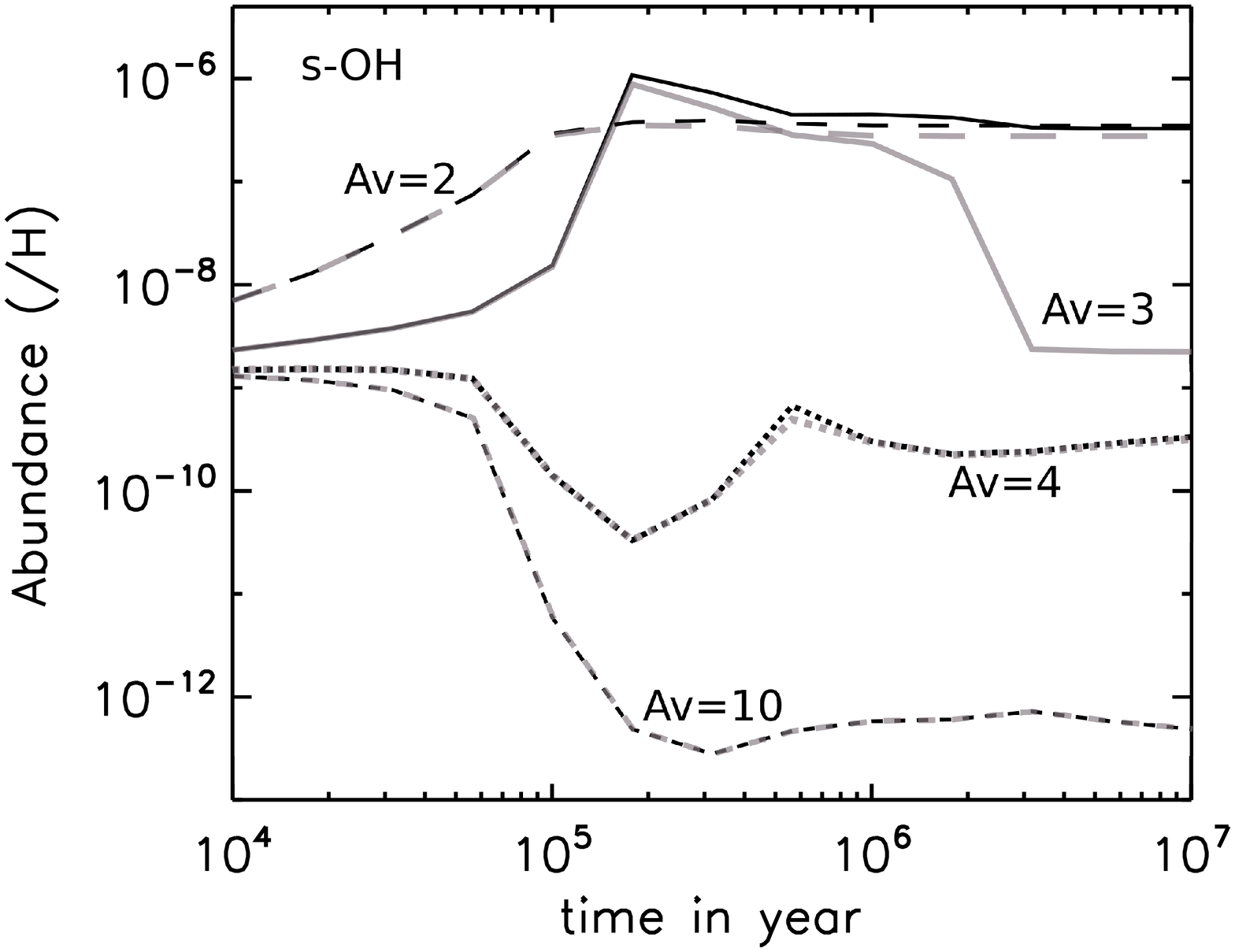}
\caption{Surface species abundances of a selection of molecules as a function of time computed for different visual extinctions. The results obtained with Model A are represented by black lines and grey lines represent those obtained with Model C. For A$_{\text{V}}$=10, the curves are superimposed.}
\label{ch3oh.eps}
  \end{center}
\end{figure}

\begin{figure}
\centering
\includegraphics[scale=0.3]{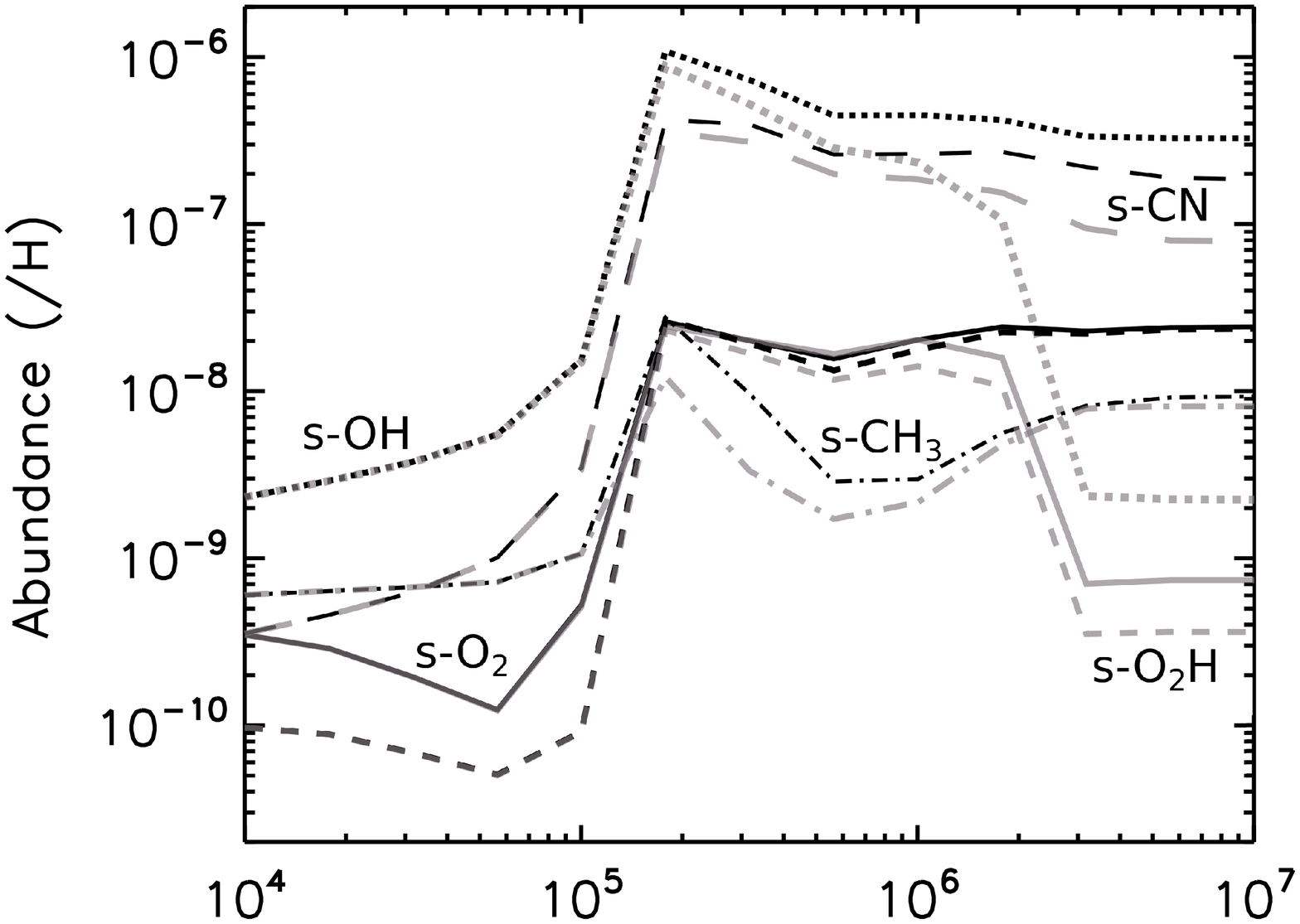}
\includegraphics[scale=0.3]{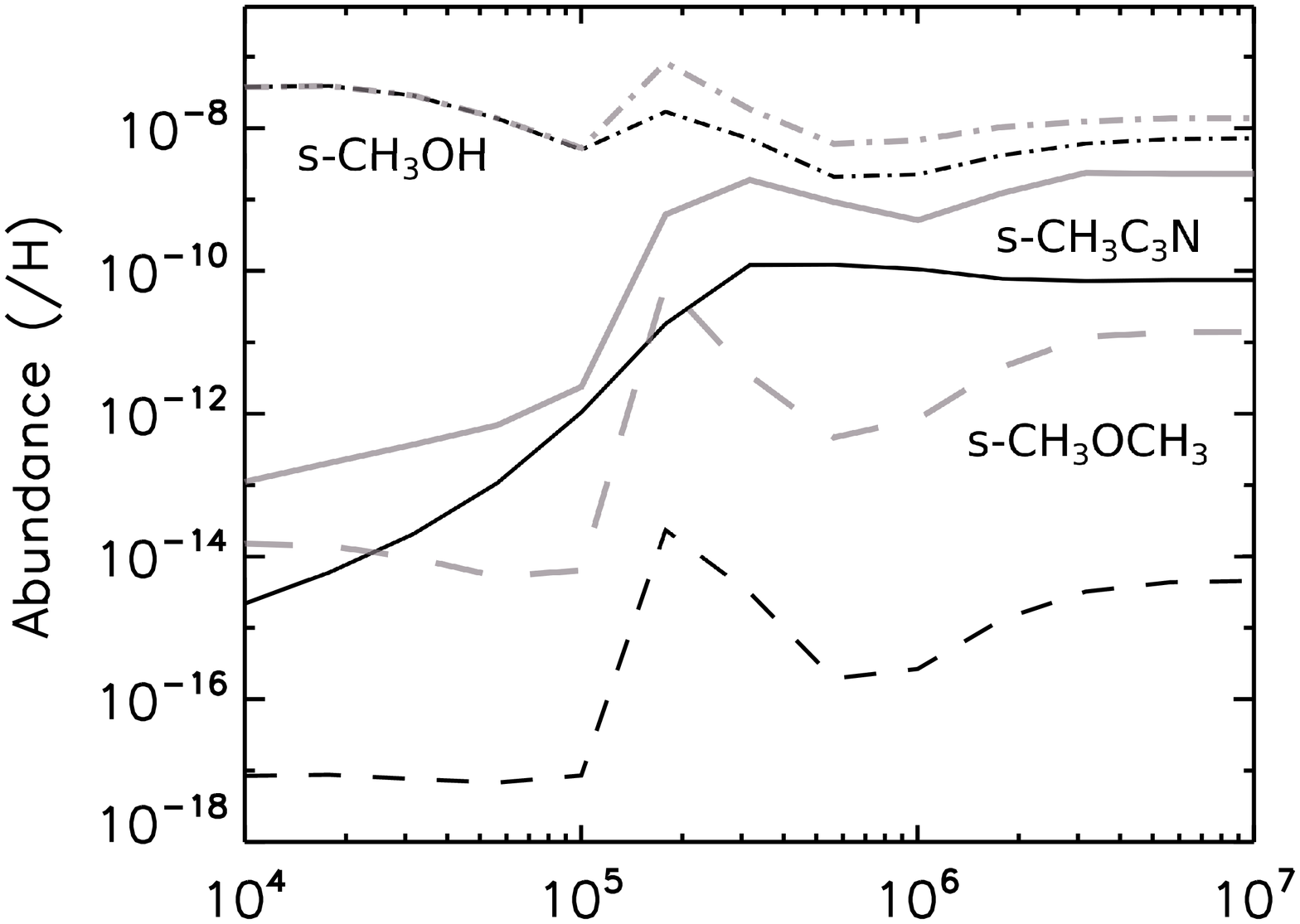}
\includegraphics[scale=0.3]{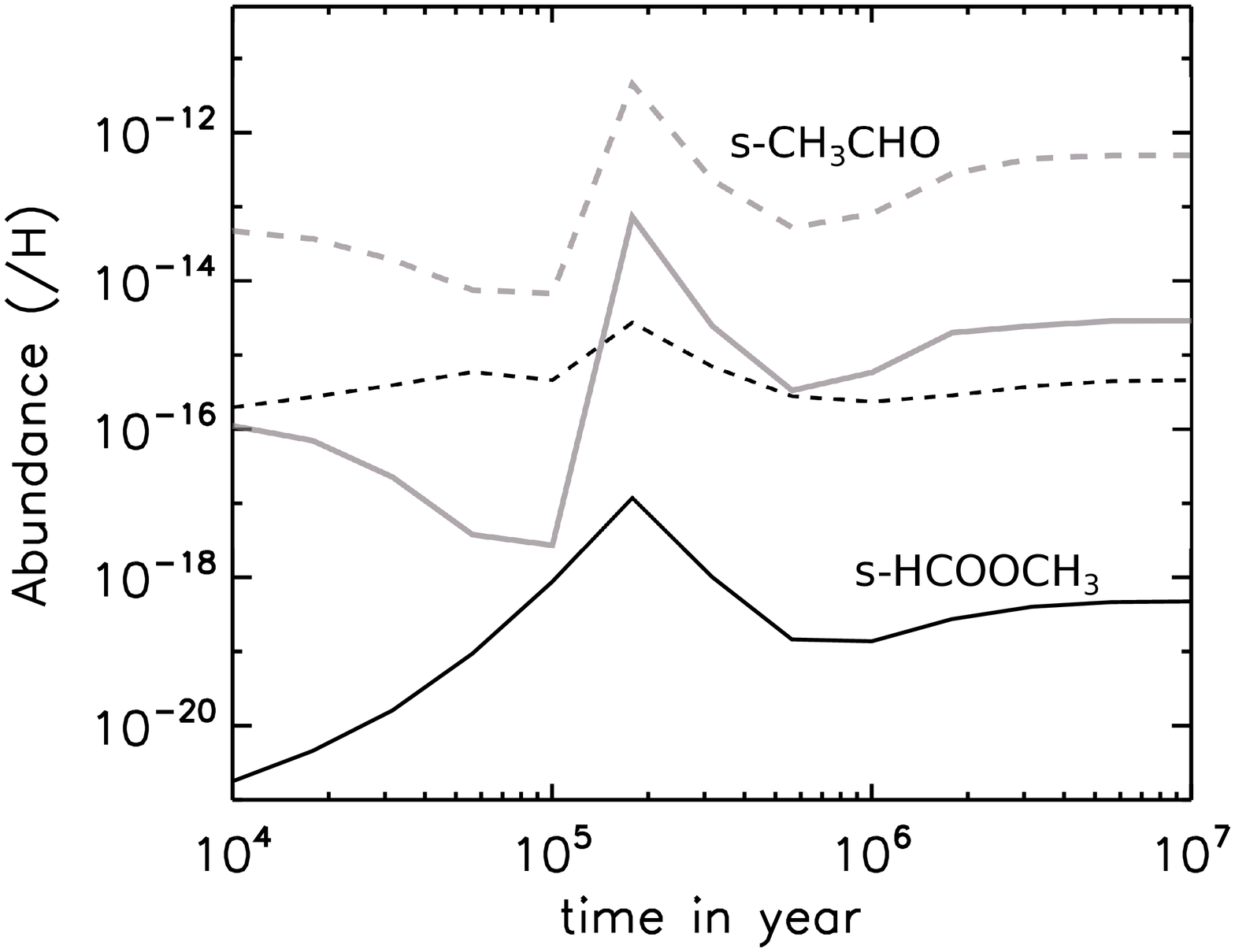}
\caption{Grain-surface abundances of simple (top) and complex molecules (bottom) as a function of time for A$_{\text{V}}$=3. Black lines are the results obtained with Model A and grey lines the results obtained with Model C.}
\label{compl.eps}
\end{figure}
We then run our models A and C for different visual extinctions (2, 3, 4 and 10) and found that the efficiency of CRID depends strongly on the visual extinction. It appears, as shown in Fig. \ref{ch3oh.eps} for three representative examples, that this mechanism has a stronger impact on the molecular abundances when the visual extinction is equal or below 3. 
Considering Model A, complex species abundances (in the gas-phase and at the surface of the grains) are decreased when the visual extinction is smaller since they are photo-dissociated by UV photons in the gas and on the grains whereas abundances of small radicals such as s-OH are increased. With CRID in Model C, the radicals at the surface of the grains are much more mobile and react to form complex molecules. In that case, even at low A$_{\text{V}}$, complex molecules such as s-CH$_{3}$CHO, s-C$_{2}$H$_{n}$, s-CH$_{3}$CH$_{2}$OH, s-CH$_{3}$OCH$_{3}$, s-CH$_{3}$COCH$_{3}$, s-CH$_{2}$OH and s-HCOOCH$_{3}$ can be formed on the surface faster than they are photo-dissociated. 
A visual extinction of 3 seems to be the value for which the mechanism is the most efficient. Above this value, the amount of radicals at the surface of the grains produced by photodissociation (by cosmic ray induced UV photons) is not large enough and below this value, it is so strong that radicals do not have enough time to recombine sufficiently on the surface of the grain. 

Figure \ref{compl.eps} shows the computed abundances of both simple and complex molecules as a function of time for Model A and Model C for an A$_{\text{V}}$ of 3. Complex surface species abundances are enhanced in Model C by a factor between $\sim$ 2 for s-CH$_{3}$OH and more than a factor 1000 for s-HCOOCH$_{3}$ at the final time, whereas abundances of simpler molecules such as s-OH, s-CN, s-O$_{2}$, s-O$_{2}$H and s-CH$_{3}$ are lowered by orders of magnitude. At higher temperature, the time for an adsorbed species to scan the entire grain being shorter, the recombination of radicals is more efficient allowing to convert simpler species into more complex molecules. Radicals abundance is then decreased while that of heavier species is increased. Gas-phase abundances are also affected as well with many complex species abundances, such as CH$_{3}$OH, CH$_{3}$CHO, HCOOCH$_{3}$, CH$_{3}$CH$_{2}$OH, CH$_{3}$OCH$_{3}$, CH$_{2}$OH and CH$_{3}$COCH$_{3}$, which are enhanced by a factor between $\sim$ 1.5 and $\sim$ 1000.
\subsubsection{Sensitivity to dust peak temperature}
\begin{figure}
\begin{center}
\includegraphics[scale=0.3]{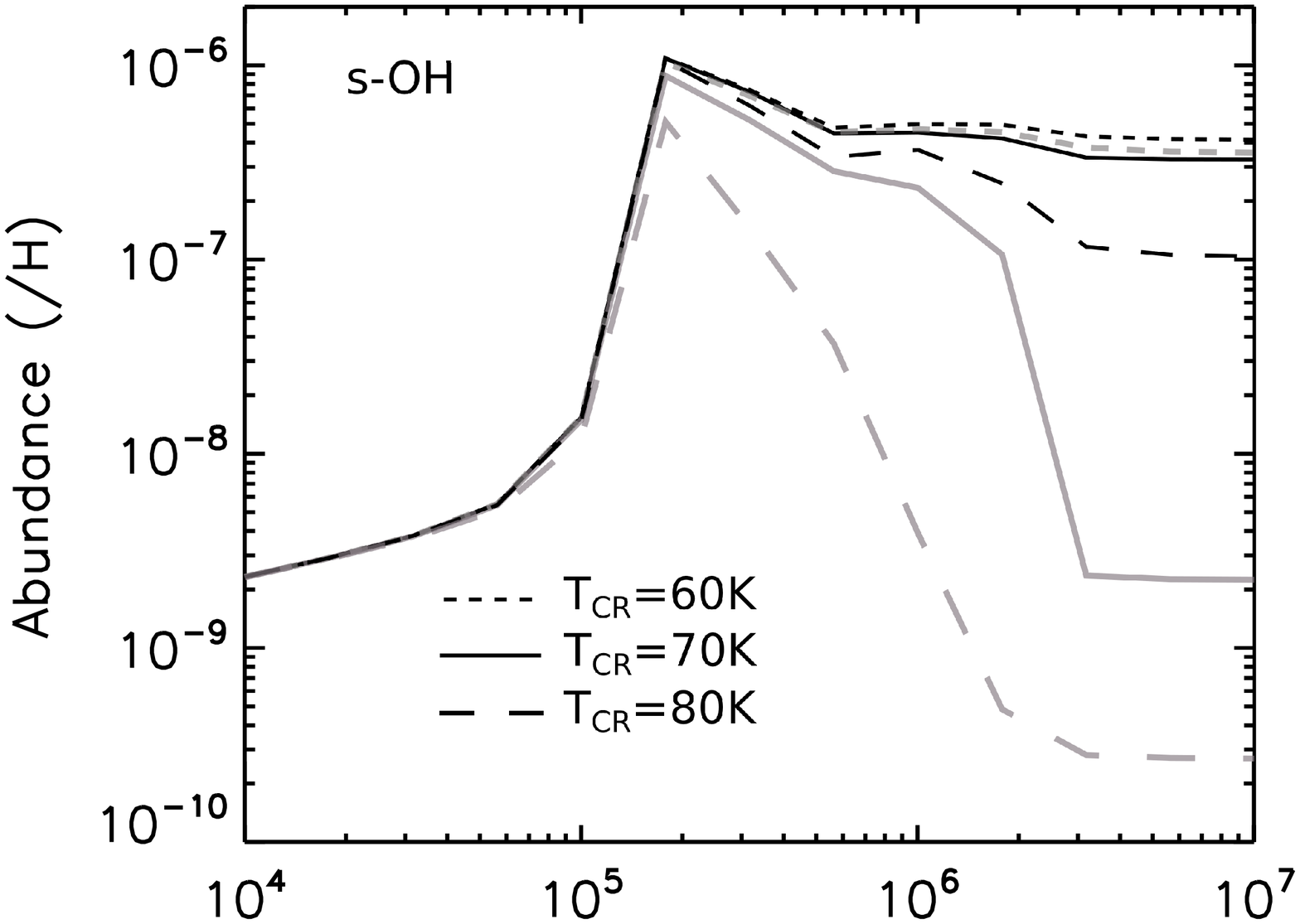}
\includegraphics[scale=0.3]{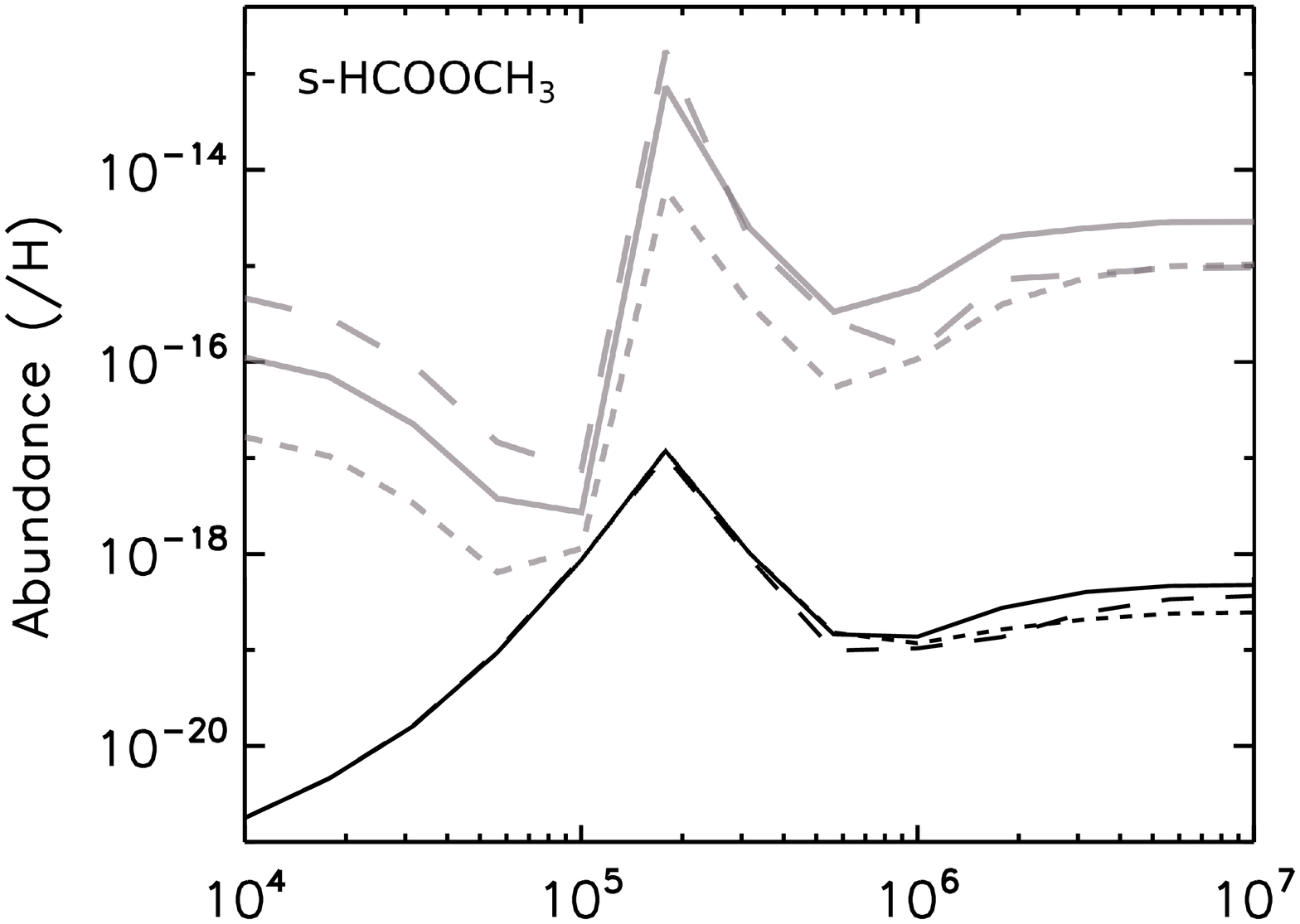}
\includegraphics[scale=0.3]{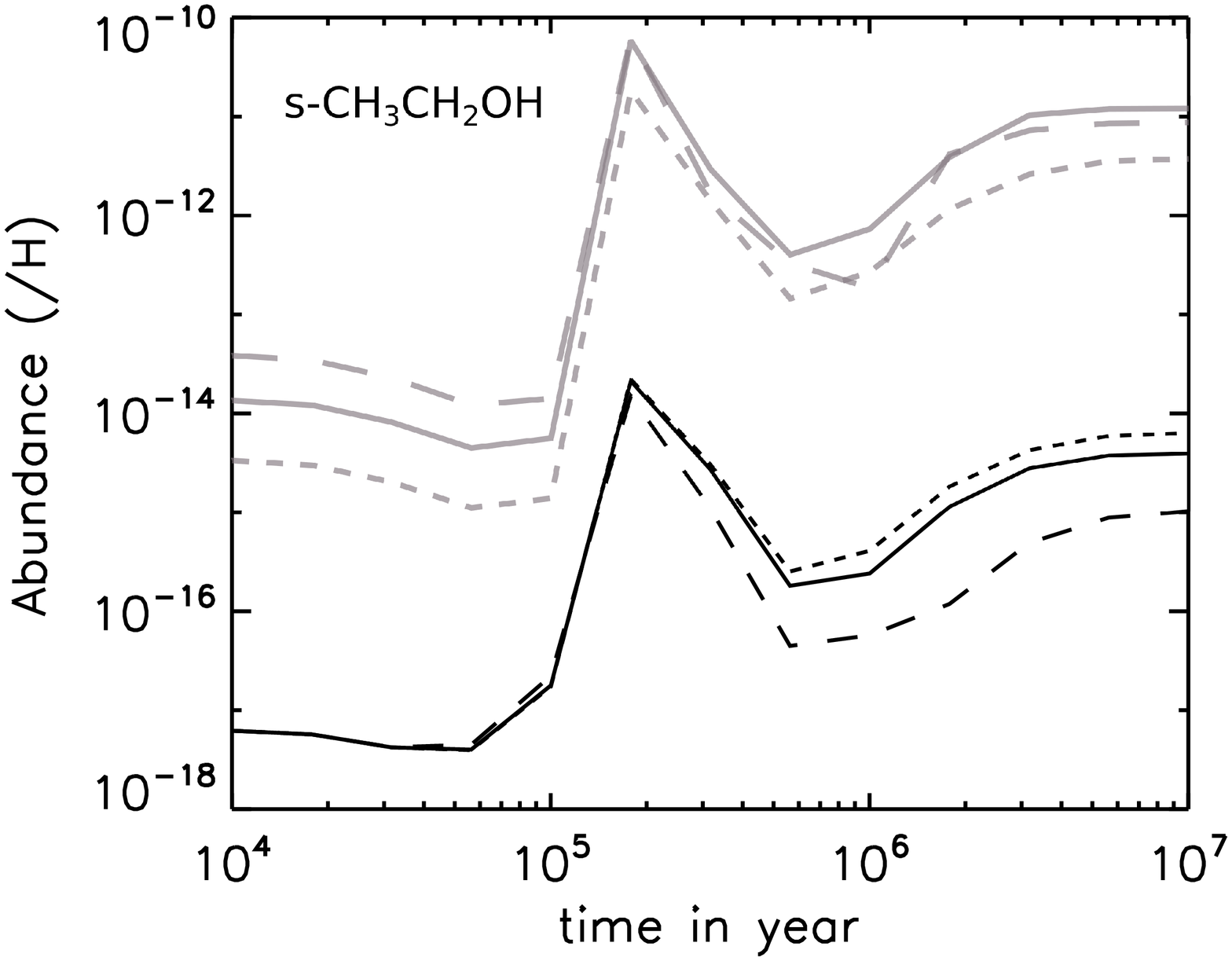}
\caption{Surface species abundances as a function of time computed for different dust peak temperatures for A$_{\text{V}}$=3. Results obtained with Model A are shown in black lines while those obtained with Model C are shown in grey lines.}
\label{temp.eps}
  \end{center}
\end{figure}
We study the effect of dust grain peak temperature induced by cosmic ray heating $T_{\text{CR}}$ (see eq. \eqref{tcr}) on the abundances. Fig. \ref{temp.eps} shows the computed surface abundances for three molecules as a function of time for three different values of $T_{\text{CR}}$: 60$\,$K, 70$\,$K (the nominal value) and 80$\,$K. Note that the value of $T_{\text{CR}}$ determines the efficiency of the CRID process but also the cosmic ray induced desorption (see section 2.2). This is why we find small differences in Model A as well. When the temperature is lower (60$\,$K), the mechanism is less efficient: radicals mobility being lower, they scan the surface more slowly and their recombination into more complex species is less efficient. On the contrary, surface radicals abundance decreases much more when the temperature is higher (80$\,$K) for two main reasons. Firstly the recombination efficiency is better due to a higher mobility and secondly, the dust temperature being warmer, species desorb back into the gas-phase more efficiently. At 80$\,$K, complex species abundance is increased between $10^{4}$ and $10^{5}$ years due to the higher mobility of simpler species. However, after $10^{5}$ years, abundance at $T_{\text{CR}}$ = 80$\,$K obtained with Model C is lower than at 70$\,$K. Indeed, both radicals and complex species desorb back into the gas-phase, the desorption coefficient rate being 10 to 10000 times higher at 80$\,$K.
\subsubsection{Sensitivity to cosmic ray ionization rate}
\begin{figure}
\begin{center}
\includegraphics[scale=0.3]{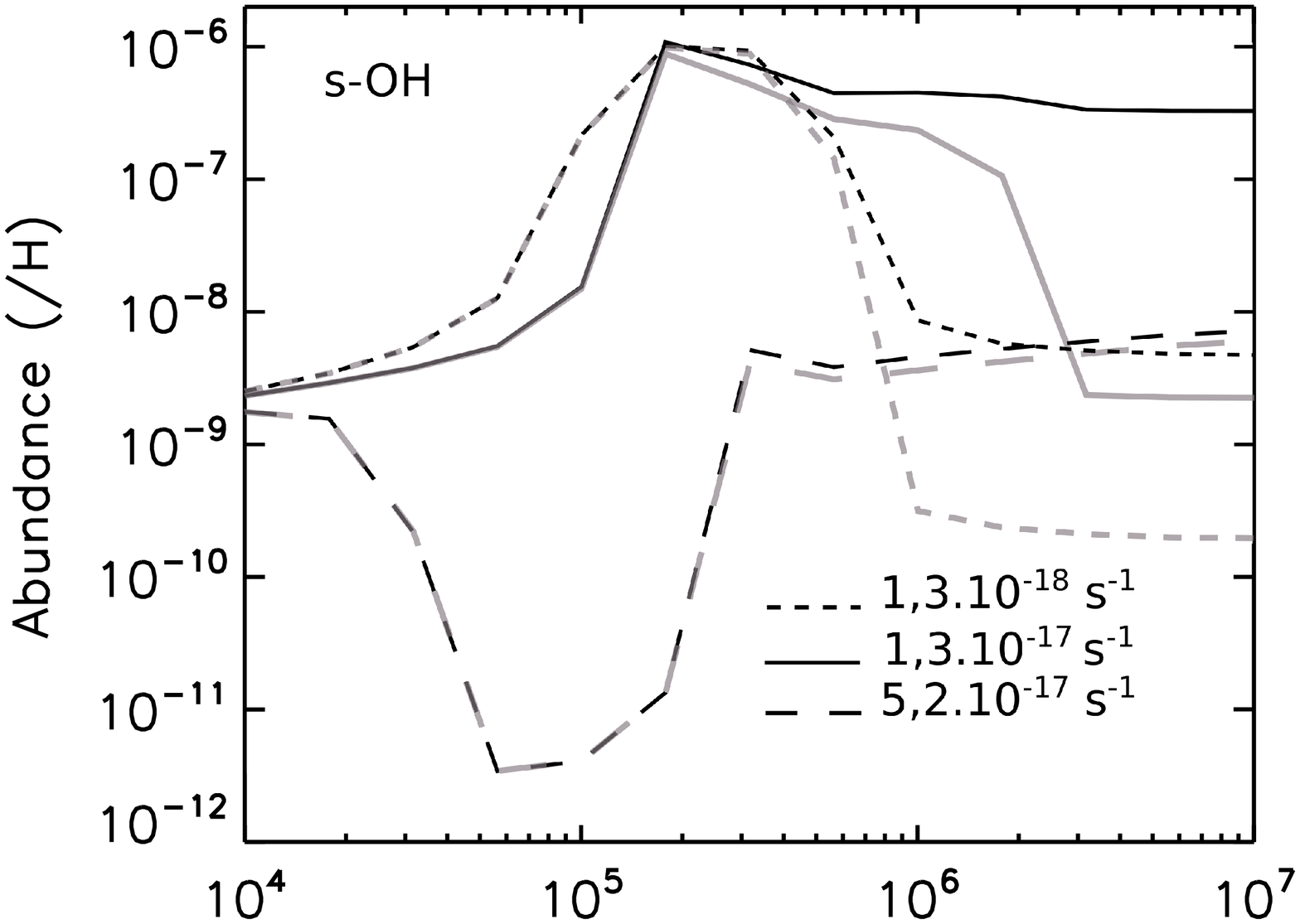}
\includegraphics[scale=0.3]{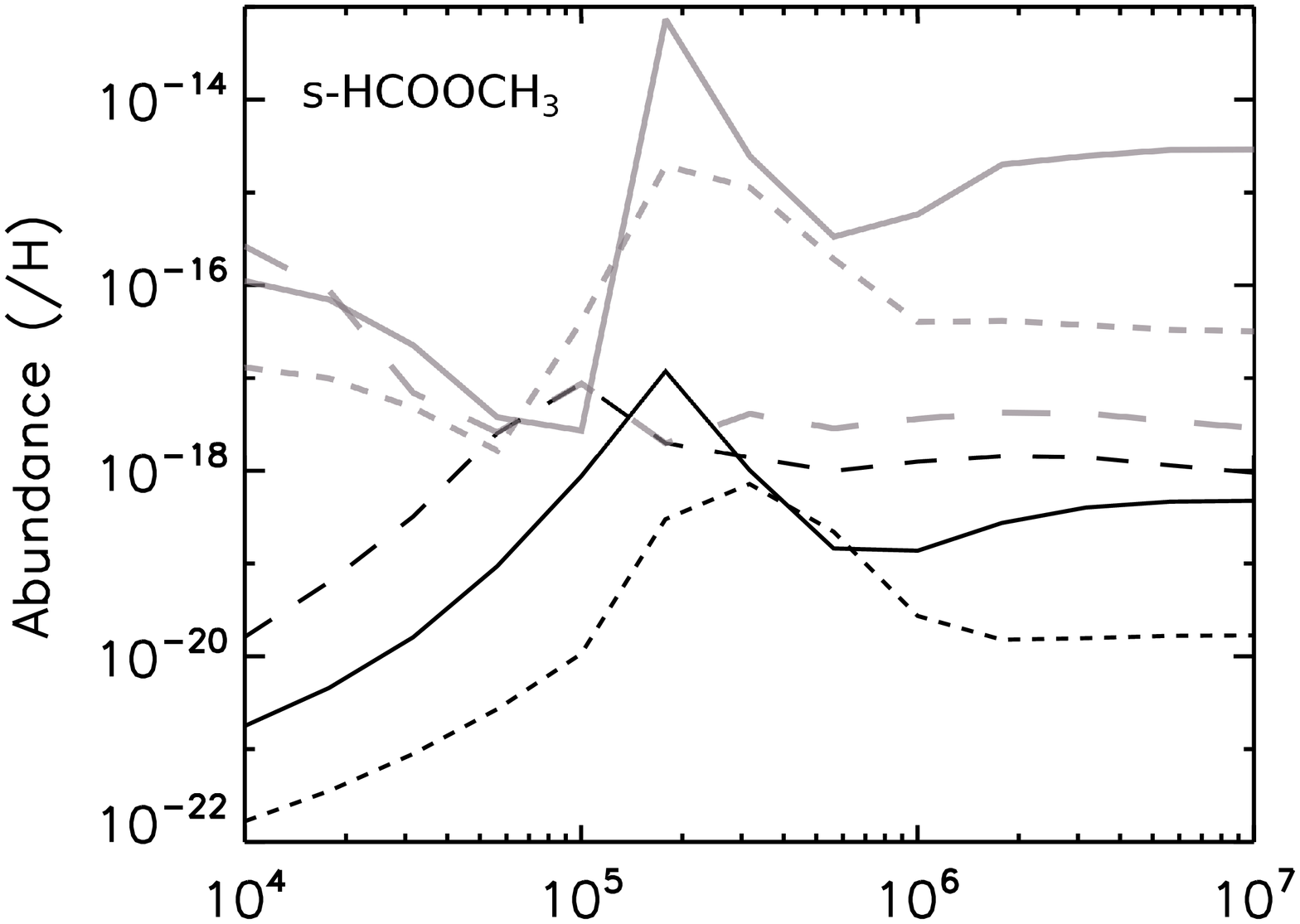}
\includegraphics[scale=0.3]{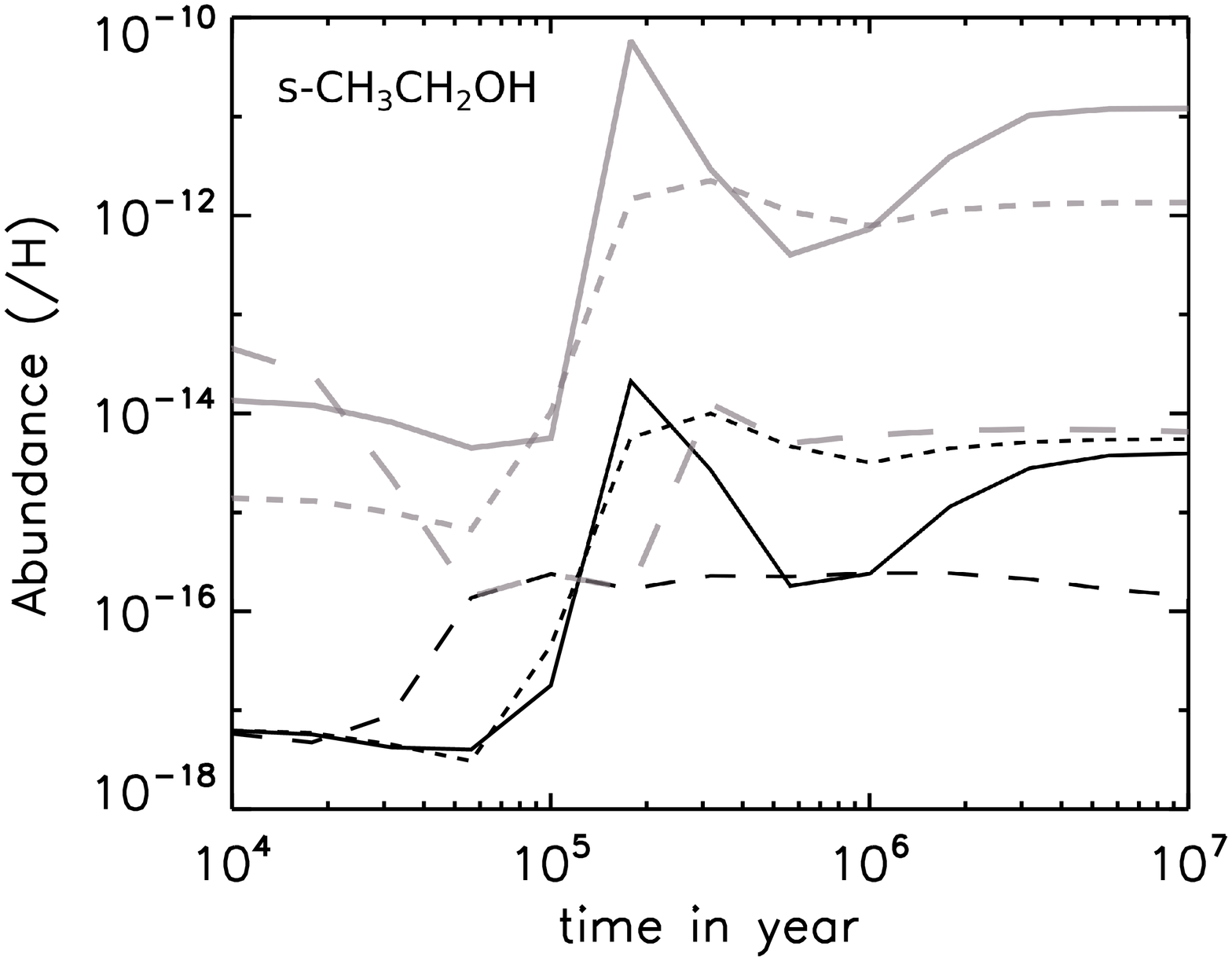}
\caption{Surface species abundances as a function of time computed for different cosmic ray ionization rates for A$_{\text{V}}$=3. Results obtained with Model A are shown in black lines while those obtained with Model C are shown in grey lines. }
\label{zetao.eps}
  \end{center}
\end{figure}
The CRID mechanism is directly proportional to the cosmic ray ionization rate $\zeta_{\text{CR}}$. Fig.~\ref{zetao.eps} shows computed abundances as a function of time for three different cosmic ray ionization rates. The solid lines represent the nominal value of the rate and the dashed lines a rate four times higher ($5.2\times10^{-17}\,$s$^{-1}$) and ten times lower ($1.3\times10^{-18}\,$s$^{-1}$) than this value. Considering the Model A only, i.e. without CRID, the effect of cosmic rays is not the same for all molecules. For many complex molecules at the surface of the grains, the time at which the abundance peaks is shifted towards longer times when $\zeta_{\text{CR}}$ is decreased. The global s-HCOOCH$_{3}$ is increased by higher $\zeta_{\text{CR}}$ whereas the abundance of s-CH$_3$CH$_2$OH is decreased. 
Using the lower $\zeta_{\text{CR}}$, the mechanism of cosmic rays impact is less important than with the nominal value, due to the fact that time spent by grains at 70$\,$K is much less important. Therefore, radicals do not have enough time to recombine in more complex species. With a rate four times higher, the efficiency is less important: the ionization rate is so strong that both simple and complex species desorb back into the gas-phase. \\ 

The CRID mechanism is much more sensitive to the visual extinction, for an A$_{\text{V}}$ too large (above 5), the dust temperature and the cosmic ray ionization rate have no impact on the abundances. 
\subsection{Combined effects of both mechanisms}
\begin{figure}
\centering
\includegraphics[scale=0.3]{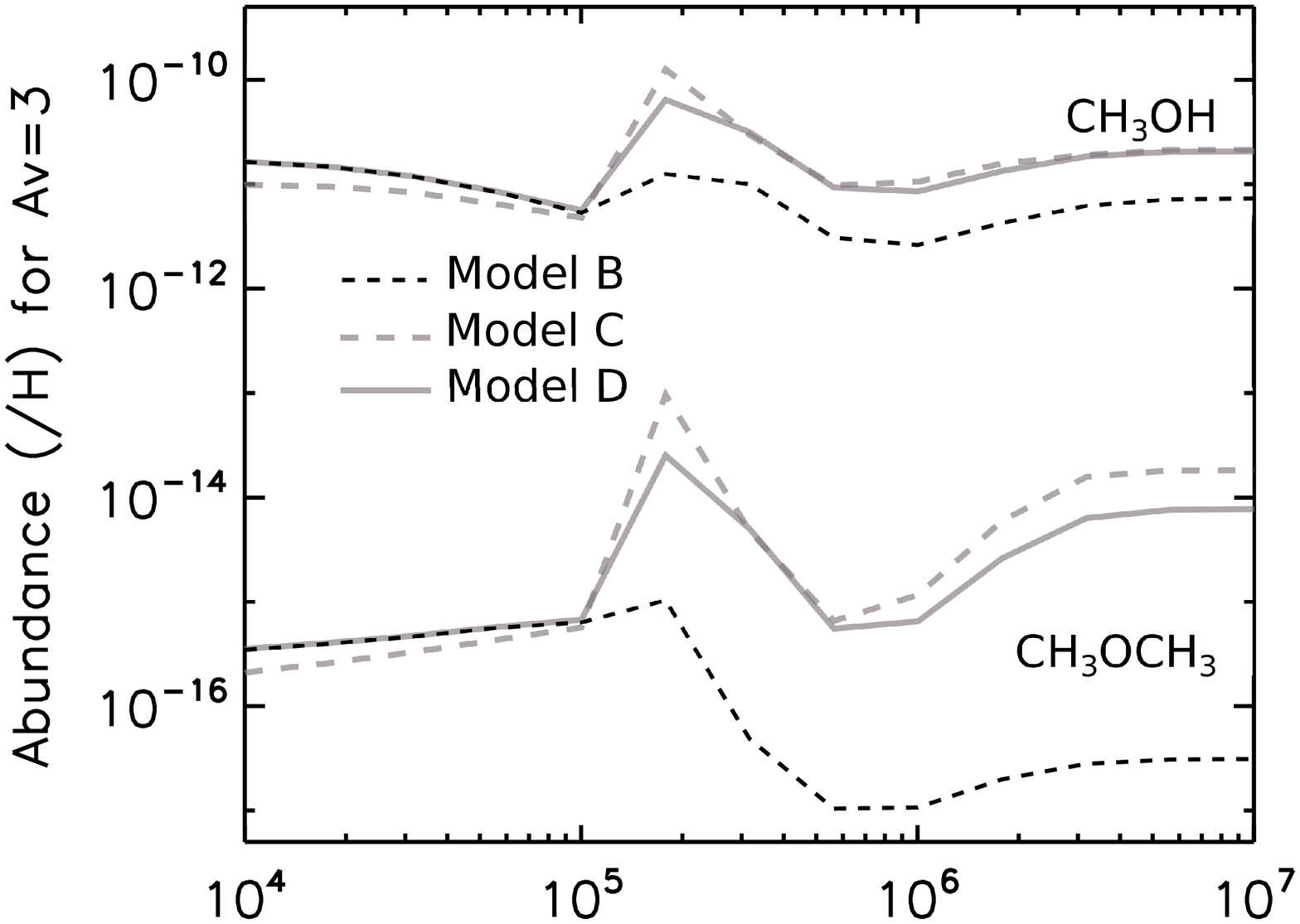}
\includegraphics[scale=0.3]{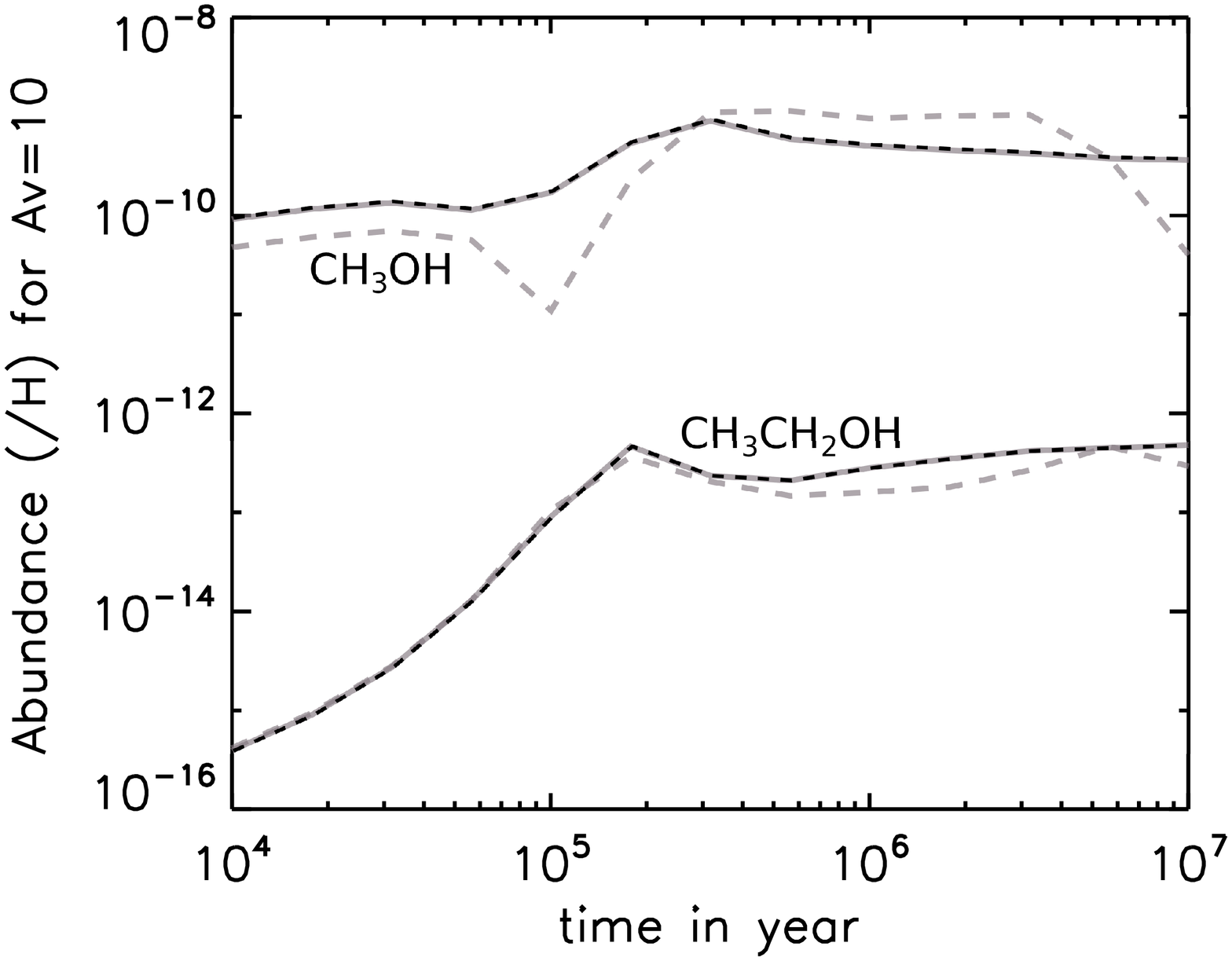}
\caption{Comparison of the three models for A$_{\text{V}}$=3 (top) and A$_{V}$=10 (bottom). For A$_{\text{V}}$=10, the curves of models B and D are superimposed.}
\label{compar.eps}
\end{figure}
We now study the effect of both mechanisms by including in the model the H, H$_2$ and O diffusion by tunneling effect and the CRID mechanism (hereafter Model D). Fig. \ref{compar.eps} shows the abundances of some molecules as a function of time computed for models B, C and D (see Table 2).
As seen before, when the value of the visual extinction is 10, CRID is inefficient and abundances are only sensitive to the quantum tunneling mechanism (Model B and Model D curves overlap). On the contrary, for a visual extinction of 3, the abundances of Model C and Model D are quite the same, suggesting that CRID dominates the diffusion of the species on the surfaces. The results which are shown in Fig. \ref{compar.eps} have been obtained for a dust temperature of 10$\,$K. At lower temperature, it is the quantum tunneling mechanism which dominates (see section 3.1.1). 
\section{Comparison with observations}
\begin{figure}
\centering
\includegraphics[scale=0.3]{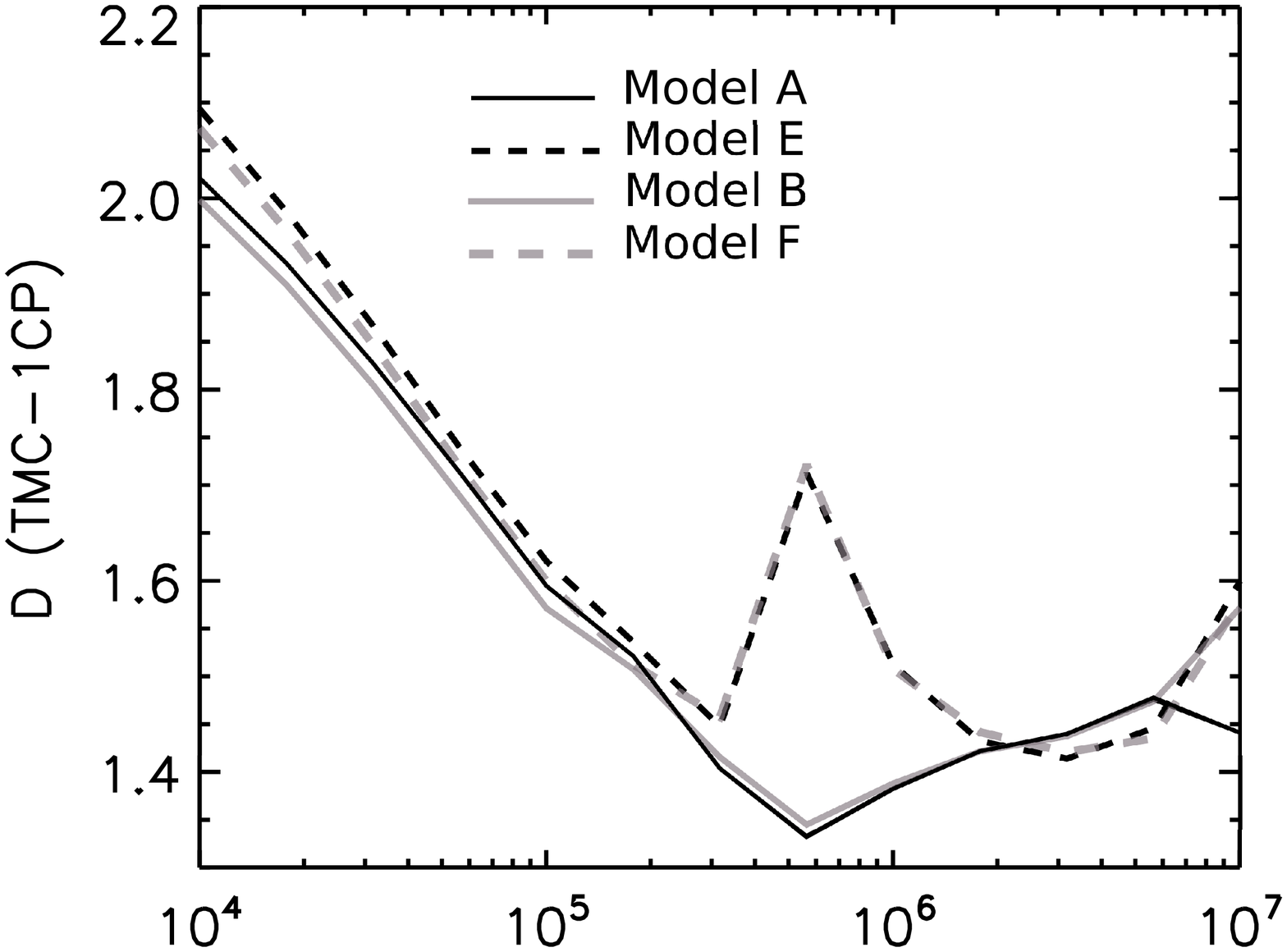}
\includegraphics[scale=0.3]{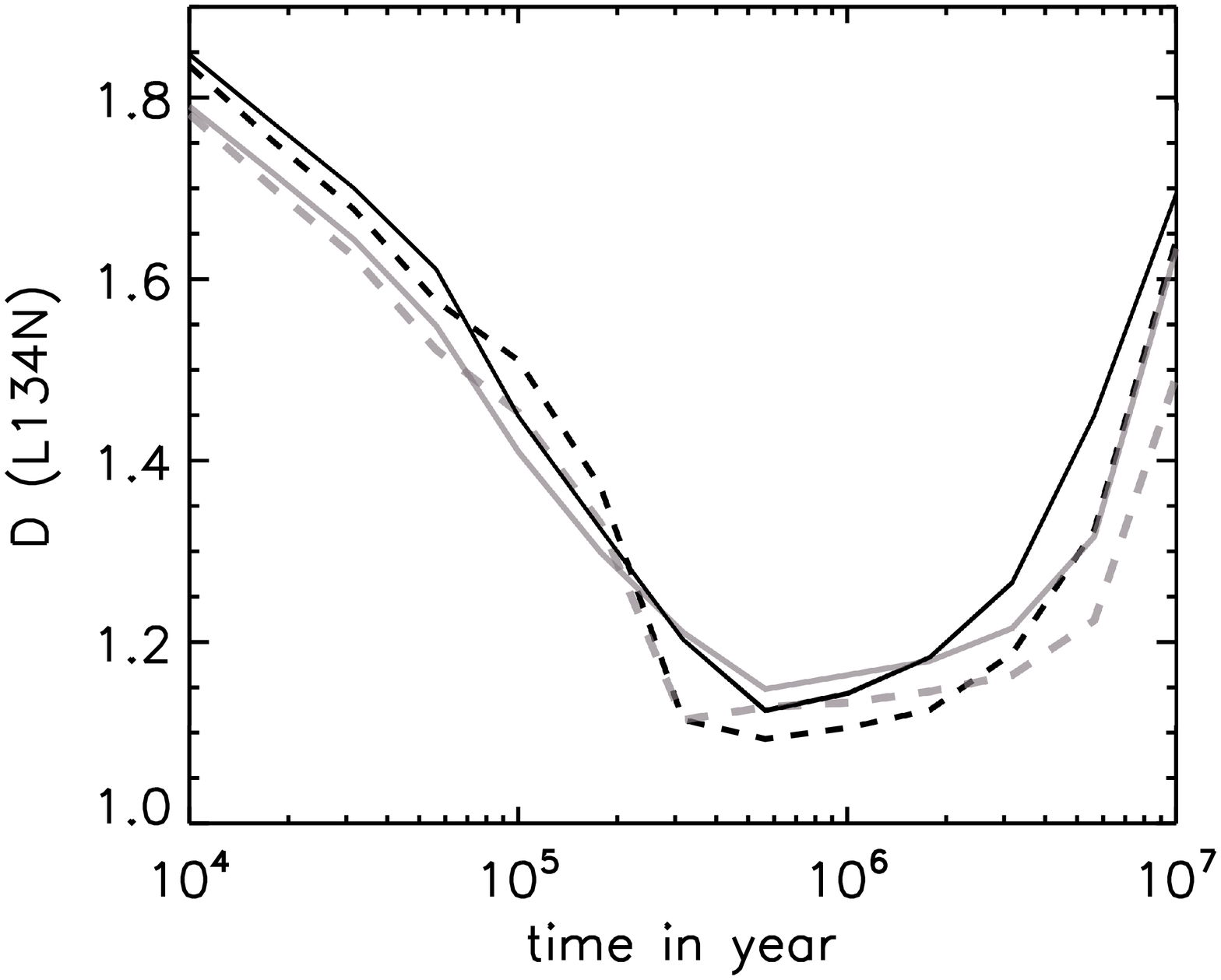}
\caption{Distance of disagreement between models predictions and observed abundances in two dark clouds : TMC-1 (CP) (top) and L134N (bottom).}
\label{conf.eps}
\end{figure}
We have compared the model predictions with some observations in two dark clouds: TMC-1 (CP: Cyanopolyynes Peak) for which we compared 51 observed species and L134N for which we compared 33 species. The observed abundances are listed in \citet{b19} (see Table 4 of their paper). Species for which only upper limits are available have been removed from this list as some of the species which are not included in our model Nautilus. To compare the model results with observations, we used the method described in \citet{b20} in which a distance of disagreement is computed for each species $i$. A distance between the models and the observations is defined as: 
\begin{equation}
D(t)=\frac{1}{N_{\text{obs}}}\sum \limits_{i} \mid\log\left[X_{i}(t)\right]-\log\left[X_{\text{obs},i}\right]\mid,
\end{equation}
where $X_{i}$ is the abundance of species $i$ computed by the model at the time $t$, $X_{\text{obs},i}$ the observed abundance of species $i$ and $N_{\text{obs}}$ is the number of species which have been observed. A smaller value of $D$ corresponds to a better agreement.\\

Figure \ref{conf.eps} presents the distance of disagreement obtained for four different models : Model A, Model B, Model E (which is the same as Model A but with a lower C/O ratio: 0.7 instead of 1.2) and Model F (which is the same as Model B but with the C/O ratio equal to 0.7). The four models have been computed for a visual extinction of 10. We do not consider Model C here since the effect of cosmic rays on the diffusion rate is quite inefficient for such value of the visual extinction. We also do not consider Model D since Model B and Model D are equivalent for A$_{\text{V}}$=10 (see section 3.3). The physical parameters are the same as our nominal cloud (see section 3).\\

The C/O ratio does not affect so much the agreement in the case of L134N. However, in TMC-1 (CP), a lower elemental O abundance matches best the observations between $2\times10^{5}$ and $1.5\times10^{6}$ years due to the higher abundance of cyanopolyynes during this range of time. 
In all cases, considering the tunnel effect for H, H$_{2}$ and O does not change significantly the agreement between the models and the observations. Those similarities can be explained by the fact that most species that present significant differences when diffusion by tunneling is included are not observed. The best agreement is obtained at $6\times10^{5}$ years for both clouds, corresponding to a time when grain-surface chemistry plays an important role.\\

In Table 4, we show the abundances predicted in the gas-phase and at the surface of the grains by Models A and B at $10^{5}$ years and $6\times10^{5}$ years for the species observed in TMC-1CP and L134N that present strong sensitivity to the diffusion by tunneling process. This represents only three species (HNCO, CH$_{2}$CHCN and CH$_{3}$OH) that are increased when diffusion by tunneling of O is considered (see Table A1, left column). At the best age previously determined ($6\times10^{5}$ years), the gas-phase abundances obtained with Model B are increased by a factor $\sim\,$1.6 and 15 for, respectively, CH$_{2}$CHCN and HNCO. However, CH$_{3}$OH gas-phase abundance is decreased by a factor $\sim\,$1.8 when the tunnel effect is considered at $6\times10^{5}$ years. The surface species abundances are also increased with Model B at $6\times10^{5}$ years for CH$_{2}$CHCN and HNCO with both a factor $\sim\,$1.4. CH$_{3}$OH grain surface abundance is decreased by a factor $\sim\,$1.5. The predicted gas-phase abundances reflect the surface ones. However, those surface abundances are much larger (100 to 1000 times larger) than the gas-phase observed ones for CH$_{3}$OH and HNCO, suggesting that other desorption mechanisms should be taken into account in order to remove those surface species back into the gas-phase and increase their gas-phase abundances.
\begin{table*}
\centering
 \begin{minipage}{25cm}
  \caption{Predicted and observed abundances of some species detected in TMC-1CP and L134N (s refers for surface species).}
  \begin{tabular}{lllllll}
  \hline \hline
  &\multicolumn{6} {l} {\bf{Abundances (/H)}}\\ \hline \hline
  \multirow{2}{*} {\bf{Molecules}} & \multicolumn{2} {l} {\bf{Model A}}  & \multicolumn{2} {l} {\bf{Model B}}  & \multicolumn{2} {l} {\bf{Observations}}\\
     &  $10^{5}\, $yrs  & $6\times10^{5}\, $yrs  & $10^{5}\, $yrs  & $6\times10^{5}\, $yrs  & {\bf{TMC-1CP} \footnote{\label{mar}\citet{b19}.}}  &{\bf{L134N}\footref{mar}} \\ \hline 
     \hline
  \multirow{2}{*} {\bf{HNCO}}& $8.8\times10^{-10}$ &$1.1\times10^{-12}$&$1.3\times10^{-9}$&$1.6\times10^{-11}$&$4\times10^{-10}$&$2.5\times10^{-10}$\\ 
  &$4.1\times10^{-7}$ (s)&$1.3\times10^{-7}$ (s)&$5.9\times10^{-7}$ (s)&$1.9\times10^{-7}$(s)&&\\ \hline
 \multirow{2}{*} {\bf{CH$_{2}$CHCN}}& $1.1\times10^{-11}$ &$1.6\times10^{-10}$&$9.2\times10^{-12}$&$2.6\times10^{-10}$&$10^{-9}$&-\\ 
   &$4.3\times10^{-13}$ (s)&$2.2\times10^{-10}$(s)&$4\times10^{-13}$ (s)&$2.8\times10^{-10}$(s)&&\\ \hline
 \multirow{2}{*} {\bf{CH$_{3}$OH}}& $1.1\times10^{-11}$ &$1.1\times10^{-9}$&$1.7\times10^{-10}$&$5.9\times10^{-10}$&$3.2\times10^{-9}$&$5.1\times10^{-9}$\\ 
   &$2.3\times10^{-7}$ (s)&$9.1\times10^{-6}$&$5.6\times10^{-7}$ (s)&$6.1\times10^{-6}$&&\\
\hline
 \hline
\label{table4}
\end{tabular}
\end{minipage}
\end{table*}

\section{Conclusions}
In this paper, we have studied the diffusion of species on the grain surfaces in molecular clouds using our gas-grain model Nautilus. In particular, we revisited the efficiency of diffusion by tunneling effect of H, H$_{2}$ and O based on recent experiments by \citet{b1}.  We also introduced a new mechanism to take into account the "boost" of surface species diffusion induced by the stochastic heating by cosmic ray particles. We call this new mechanism Cosmic Rays Induced Diffusion (CRID). Species most affected by those two mechanisms are listed in Table A1.

The diffusion by tunnel effect allows surface radicals to move faster at lower temperature which increased the abundances of simple and complex species (in the gas-phase and at the surface of the grains), such as HOOH, CH$_{3}$OCH$_{3}$, for typical dense cloud physical parameters and ages. This effect becomes even more important for temperatures below 10$\,$K. The newly introduced CRID mechanism increases the mobility of radicals at the surface of the grains, which can recombine with each other and form more complex molecules. However, we notice that this mechanism is only efficient when the visual extinction is smaller or equal to 3 for 0.1$\,$$\mu$m grains. For such values of visual extinction, we show that complex species abundances can be increased significantly while those of simpler molecules are decreased. This is because the photodissociation of molecules at the surface of the grains produces more radicals that can then react and their formation rate is then larger than their destruction one.
 
The general agreement of our chemical model with molecular abundances observed in the two clouds TMC-1CP and L134N is not significantly changed by considering the diffusion by tunneling of H, H$_{2}$ and O because most of the observed species are not affected by this process. Only three are significantly sensitive: HNCO, CH$_{2}$CHCN and CH$_{3}$OH. For those species, the predicted gas-phase abundances are larger and come closer to the observations. Furthermore, on the surface, their abundances are much larger than the gas-phase observed ones for CH$_{3}$OH and HNCO. For the complex species observed in pre-stellar cores by \citet{b24} and \citet{b25}, the diffusion of atomic oxygen by tunneling is not important enough to reproduce the large gas-phase abundances for HCOOCH$_{3}$, CH$_{3}$OCH$_{3}$, CH$_{2}$CO and CH$_{3}$O. Even on the surfaces, their abundances stay at a level lower than the gas-phase observed ones.
\section*{Acknowledgments}
The authors thank the French CNRS/INSU program PCMI for their partial support of this work. VW and FH research are funded by the ERC Starting Grant (3DICE, grant agreement 336474). The authors also thank the referee for suggestions which helped to improve this paper.

\newpage
\appendix
\section[]{}
\begin{table}
 \begin{minipage}{7.5cm}
 \caption{Species for which abundances (above $10^{-13}$ (/H)) are affected by more than a factor 2 (with respect to Model A) at $10^{6}$ years. s- refers for surface species. Species detected towards TMC-1 (CP) are marked with $^{*}$, those detected towards  L134N are marked with $^{\star}$.}
\begin{tabular}{lllll}
\hline
\bf Model B (for A$_{\text{V}}$=10)& \bf Model C (for A$_{\text{V}}$=3)\\
\hline
C$_{2}$H$_{4}$&H$_{2}$O\\
CH$_{2}$CCH&OH$^{*\star}$\\
CH$_{3}$OH$^{*\star}$&CO$_{2}$\\
NH$_{2}$CHO&SiO$_{2}$\\
CH$_{2}$CHCN$^{*}$&HCOOH$^{\star}$\\
HCOOCH$_{3}$&CH$_{3}$OH$^{*}$\\
CH$_{3}$CH$_{2}$OH&C$_{2}$H$_{n}$ (n = 5; 6)\\
CH$_{3}$OCH$_{3}$&O$_{3}$\\
O$_{3}$&CH$_{2}$OH\\
HNCO$^{*\star}$&s-C$_{2}$H$_{n}$ (n = 5; 6)\\
HC$_{2}$O&s-CH$_{2}$OH\\
CH$_{2}$OH&s-CH$_{3}$CH$_{2}$OH\\
s-C$_{2}$H$_{n}$ (n = 4; 5)&s-CH$_{2}$\\
s-CH$_{3}$CH$_{2}$OH&s-H$_{2}$CCN\\
s-CCS&s-CH$_{3}$C$_{n}$N (n = 1; 3; 5; 7)\\
s-CH$_{2}$CCH&s-CH$_{3}$OCH$_{3}$\\
s-CH$_{2}$CHCN&s-CH$_{3}$OH\\
s-C$_{3}$O&s-CH$_{3}$NH$_{2}$\\
s-CH$_{2}$&s-CN\\
s-H$_{2}$CCO&s-CO$_{2}$\\
s-CH$_{3}$&s-CS\\
s-CH$_{3}$OH&s-H$_{2}$CO\\
s-CN&s-H$_{2}$CS\\
s-CO&s-H$_{2}$O\\
s-CO$_{2}$&s-HCOOH\\
s-H&s-HNO\\
s-H$_{2}$CO&s-NH$_{2}$CHO\\
s-H$_{2}$CS&s-NS\\
s-HCOOCH$_{3}$&s-O\\
s-HNCO&s-O$_{2}$H\\
s-HNO&s-O$_{3}$\\
s-C$_{n}$H$_{2}$ (n = 4; 5; 6; 8; 9) & s-OCS\\
s-NH&s-OH\\
s-NH$_{2}$&s-S\\
s-NH$_{2}$CHO&s-SO$_{2}$\\
s-NO&-\\
s-OH&-\\
s-SO&-\\
s-CS&-\\
s-HOOH&-\\
s-H$_{2}$S&-\\
s-NS&-\\
s-O$_{2}$&-\\
\hline
 \end{tabular}
 \end{minipage}
 \end{table}

\bsp

\label{lastpage}

\end{document}